# Generalized reduced-order particle-in-cell scheme for Hall thruster modeling: concept and in-depth verification in the axial-azimuthal configuration

Maryam Reza [1†], Farbod Faraji[1], Aaron Knoll[1]

[1]Plasma Propulsion Laboratory, Department of Aeronautics, Imperial College London, Exhibition Road, London, SW7 2AZ, United Kingdom

**Abstract**:

Reduced-order particle-in-cell (PIC) scheme is a novel modeling approach that enables computationally efficient electrostatic kinetic simulations of plasma. Through our previous publications, we demonstrated the potentials of a preliminary implementation of this novel PIC scheme to resolve the multi-dimensional plasma processes and their interactions in a Hall thruster in a manner close to traditional multi-dimensional electrostatic PIC simulations. In this work, we first introduce a mathematically mature formulation for the dimensionality reduction of Poisson's equation, which enables the generalized reduced-order "quasi-multi-dimensional" PIC scheme. We present the proof that the dimensionally reduced Poisson's equation converges to the multi-dimensional one in the limit. A reduced-dimension Poisson solver that incorporates the dimensionality reduction formulation is then verified for general 2D Poisson problems. Next, we present the results of several quasi-2D axial-azimuthal kinetic simulations we have performed using the fully generalizable implementation of the reduced-order electrostatic PIC scheme. Based on the detailed studies and analyses reported in this article, we show that the recent improvements in the formulation of the reduced-order PIC scheme notably augments its predictive capability, enabling the approach to reproduce quite accurately the results from reference full-2D axial-azimuthal PIC simulations in several conditions using very low-order approximations of the 2D problem. Moreover, higher order approximations allow us to recover the same characteristics, behaviors and effects reported in the literature regarding the azimuthal instabilities in Hall thrusters with simulations that still offer several factors reduction in the computational cost compared to traditional 2D3V PIC simulations. Finally, we discuss a series of sensitivity analysis results, including the influence of the cathode boundary condition and the azimuthal domain length on the predictions of the quasi-2D simulations. Accordingly, we have outlined the ongoing activities on the reduced-order scheme in the conclusions, which are aimed at gradually paving the way toward cost-effective and comprehensive three-dimensional studies of the physics in Hall thrusters.

## Section 1: Introduction

Hall thrusters are plasma-based in-space propulsion devices that can be used in a variety of mission scenarios, ranging from on-orbit servicing to near-Earth transportation and interplanetary exploration. They rely on electromagnetic fields to ionize the propellant and to accelerate the created plasma to generate thrust. Although Hall thrusters are relatively simple from an engineering standpoint and their basic operating principle is quite straightforward to understand, their underlying physics of operation is rather intricate. The complexity arises, in part, from the fact that, in the presence of a perpendicular configuration of electric and magnetic field in these thrusters, the plasma is subject to strong gradients and anisotropies, which are sources of a variety of instabilities and turbulence across a vast range of spatial and temporal scales. From an operational perspective, these instabilities significantly affect the thruster's performance and lifetime, which ties the understanding of the physics to the development of efficient devices and their reliable in-space application.

Accordingly, Hall thrusters have gathered a notable scientific attention in the past decades. A considerable portion of research on Hall thrusters has been focused on deriving insights from the investigation of plasma processes and interactions using high-fidelity kinetic simulations to realize predictive models of Hall thrusters in a bid to enable their cost-effective numerically aided design and testing [1][2]. However, despite numerous efforts to this end [3][4][5][6][7], reliable prediction of the plasma behavior in Hall thrusters is still an elusive target. This is, in part, due to the fact that the dynamics of the charged particles and, in particular, the electrons' cross-field mobility in the presence of plasma instabilities and turbulence is not yet fully understood.

[†]Corresponding Author (m.reza20@imperial.ac.uk)

In this regard, it is becoming increasingly evident that a comprehensive understanding of the three-dimensional kinetic nature of the plasma dynamics in a Hall thruster is essential to fully comprehend the interplay among various plasma phenomena in the thruster and to establish links between these phenomena and their effect on the global discharge behavior [8]. Achieving this goal requires using 3D kinetic simulations since important aspects of the underlying physics can be lost by resorting to lower-dimensional 2D kinetic codes. However, 3D kinetic simulations are currently computationally unfeasible for real-world thruster geometries. Thus, in the absence of a comprehensive knowledge regarding the multi-scale, multi-dimensional plasma phenomena and their complex interactions in Hall thrusters, developing cost-effective predictive numerical simulations is yet to be achieved.

In this respect, noting the necessity of 3D kinetic simulations to thoroughly study plasma dynamics in Hall thrusters, and to address the issue of enormous computational cost associated with conventional multi-dimensional kinetic simulations that renders them impractical for real-size Hall devices [9], we introduced the reduced-order "pseudo-2D" electrostatic particle-in-cell (PIC) scheme in a prior publication [10]. In essence, our reduced-order PIC scheme relies on a dimensionality-reduction technique to decompose the multi-dimensional Poisson's equation in the Vlasov-Poisson system into several one-dimensional components along the involved simulation directions. This, in turn, remarkably reduces the required number of computational cells and the total number of macroparticles.

In our previous works [10][11], we presented a preliminary formulation to split the 2D Poisson's equation into a set of decoupled 1D Ordinary Differential Equations (ODE), which yielded the "pseudo-2D" PIC scheme. We studied the feasibility of the concept of reduced-order PIC scheme through performing several *pseudo-2D* kinetic simulations along various coordinates of a Hall thruster, i.e., axial-azimuthal, azimuthal-radial and axial-radial, comparing the results against the observations in full-2D reference cases. In this respect, although the feasibility studies were conducted with simulations that relied on a rather simple formulation for the splitting of Poisson's equation, the essential capability of the approach to resolve the main physics and the interactions among plasma phenomena in different 2D configurations was demonstrated.

According to the remarkable potential observed for the preliminary pseudo-2D scheme, we present in this article a mathematically mature formulation for the decomposition of Poisson's equation that leads to a generalizable dimensionality-reduction technique. Incorporating this dimensionality-reduction formulation in the PIC code yields the generalized implementation of the reduced-order or "quasi-multi-dimensional" PIC scheme. This novel PIC scheme allows us to ultimately realize computationally affordable high-fidelity 3D kinetic simulations of Hall thrusters.

Nevertheless, before proceeding to 3D simulations, the novelty of our approach necessitates following a meticulous verification strategy. Indeed, in the absence of 3D benchmarks, our strategy to verification of the reduced-order PIC scheme involves carrying out quasi-2D PIC simulations in various 2D configurations of a Hall thruster and to compare our results against those from full-2D PIC simulations. This strategy enables us to differentiate the dominant phenomena along different thruster's dimensions and their impact(s) on the plasma processes along the other dimensions to assess whether the captured physics adequately resembles that represented in full-2D simulations. Based on this verification framework, the focus of the present article is to report in-depth verification results of the generalized reduced-order PIC in the 2D axial-azimuthal coordinates of a Hall thruster.

To this end, the remainder of this article is organized as follows: in Section 2, we present the mature formulation for the dimensionality reduction of Poisson's equation, discuss some numerical implementation details associated with the reduced-order electrostatic PIC scheme, and provide an overview of the numerical advantages that this scheme has over traditional multi-dimensional PIC codes. In section 3, we demonstrate the mathematical validity of the dimensionality-reduction formulation. In this respect, we first provide the mathematical proof that the system of equations obtained by applying the dimensionality-reduction technique converges in the limit to the multi-dimensional Poisson's equation. Second, we show the numerical verifications of the dimensionality-reduction approach for several canonical 2D Poisson problems. Section 4 is dedicated to the discussion of the results from the quasi-2D axial-azimuthal PIC simulations using the "low-order" approximations of the 2D problem. This is to verify that the "low-order" reduced-order PIC simulations allow us to recover with reasonable accuracy similar results and observations to those from full-2D reference simulations in the literature. We begin Section 4 by presenting the setup of the quasi-2D axial-azimuthal simulations. Next, we discuss the predictions of "low-order" quasi-2D simulations in various simulation conditions in terms of the time-averaged distribution of plasma properties, the characteristics of the resolved azimuthal instabilities and their contribution to electrons' cross-field transport. In Section 5, we present a set of results from "high-order" quasi-2D simulations to demonstrate that our reduced-order PIC simulation results practically converge to those from full-2D simulations, capturing the same physics at a reduced computational cost. We then discuss in Section 6 the sensitivity of the reduced-order axial-azimuthal simulations to various numerical aspects, including the implementation of the cathode boundary condition, the azimuthal length of the domain, and the number of macroparticles per cell in both the "low"- and "high-order" approximation limits. Finally, Section 7 provides the conclusions from the present study and outlines the future work.

## Section 2: The concept of the reduced-order PIC scheme

As mentioned in Section 1, the reduced-order PIC scheme is first and foremost enabled through the dimensionality reduction of the multi-dimensional Poisson's equation. Accordingly, we first present in this section the mature formulation of the dimensionality-reduction approach. Second, we discuss the main algorithmic modifications that are made in the PIC code as a result of the integration of the dimensionality-reduction approach to yield the reduced-order PIC scheme. We also present the computational advantage of the reduced-order PIC compared to the conventional schemes.

Without losing generality, and in line with the objective of this work to verify the reduced-order PIC scheme in a 2D axial-azimuthal configuration, we focus from this point forward on a two-dimensional problem and, thus, the 2D Poisson's equation.

### 2.1. Formulation of the dimensionality-reduction approach for Poisson's equation

To introduce the basic concepts behind the dimensionality-reduction approach, we first refer to Figure 1 in which a 2D domain with the extents $L_x$ and $L_y$ along the x- and y-direction, respectively, is represented by two perpendicular computation grids, one along x and one along y. The computational cells along the x-direction have the width $\Delta x$, determined according to the physics of the problem, and are extended over the entire $L_y$ extent of the domain. Similarly, the computational cells along y have the width $\Delta y$ and a length equal to $L_x$. We term this representation of the domain the "single-region" decomposition, since we have a single set of perpendicular grids representing the entire domain.

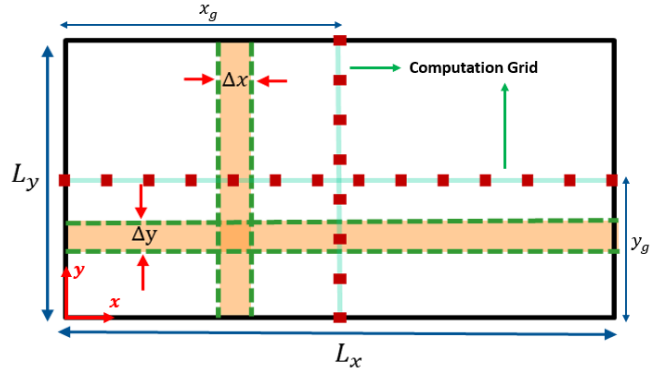

Figure 1: Schematic of the computational domain and the defined "Region" for the single-region decomposition

We define two potential functions $\phi^x$ and $\phi^y$, which are, by definition, only a function of the x- and y-coordinate, respectively, within a region. We approximate the 2D potential function $\phi(x,y)$ as the superimposition of these two potential functions, i.e.,

$$\phi(x,y) = \phi^x(x) + \phi^y(y). \tag{Eq. 1}$$

Substituting Eq. 1 in the 2D Poisson's equation

$$\nabla^2 \phi(x,y) = \frac{\partial^2 \phi(x,y)}{\partial x^2} + \frac{\partial^2 \phi(x,y)}{\partial y^2} = -\frac{\rho(x,y)}{\epsilon_0}, \tag{Eq. 2}$$

and integrating once along the horizontal (x) and once along the vertical (y) computation grid yields a coupled system of 1D Ordinary Differential Equations for $\phi^x$ and $\phi^y$, respectively,

$$\frac{d^2\phi^x(x)}{dx^2}L_y + \frac{d\phi^y(y)}{dy}\bigg|_{y_g+\frac{L_y}{2}} - \frac{d\phi^y(y)}{dy}\bigg|_{y_g-\frac{L_y}{2}} = -\frac{1}{\epsilon_0}\int_{y_g-\frac{L_y}{2}}^{y_g+\frac{L_y}{2}} \rho(x,y)dy, \tag{Eq. 3}$$

$$\frac{d^2\phi^y(y)}{dy^2}L_x + \frac{d\phi^x(x)}{dx}\bigg|_{x_g+\frac{L_x}{2}} - \frac{d\phi^x(x)}{dx}\bigg|_{x_g-\frac{L_x}{2}} = -\frac{1}{\epsilon_0}\int_{x_g-\frac{L_x}{2}}^{x_g+\frac{L_x}{2}} \rho(x,y)dx. \quad \text{(Eq. 4)}$$

In the above equations, $\epsilon_0$ is the permittivity of free space, and $x_g$ and $y_g$ are the locations of the y and x computation grid, respectively, with respect to the origin. $\rho(x,y)$ is the 2D charge density distribution.

Clearly, the above single-region decomposition has a limited applicability in terms of the solutions it can reproduce. In order for the dimensionality-reduction approach and, hence, the reduced-order PIC scheme to be readily generalizable, we, thus, extend the single-region decomposition formulation described above to a "multi-region" decomposition (Figure 2), where the 2D domain is decomposed into a grid of $N \times M$ regions, denoted by $\Omega_{n,m}$. The regions' boundary is defined such that it is equidistant from the two computational nodes adjacent to it. *Within each region*, there is a set of perpendicular grids along which the 1D potential functions $\phi_n^x$ and $\phi_m^y$ are resolved. Moreover, the 2D potential function is now approximated as

$$\phi(x,y) = \phi_n^x + \phi_m^y, \quad \text{with} \quad x, y \in \Omega_{n,m}. \quad \text{(Eq. 5)}$$

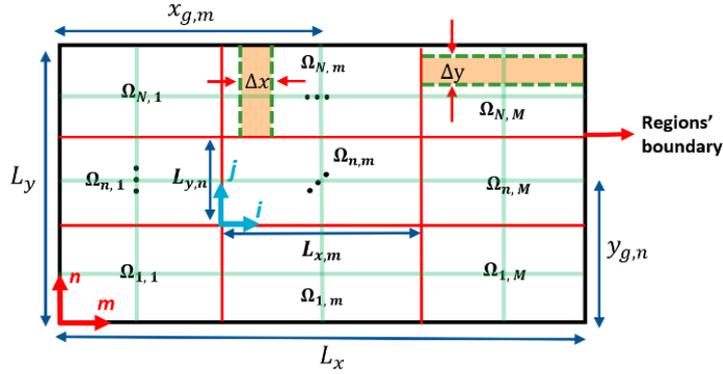

Figure 2: Schematic of the "multi-region" domain decomposition for the dimensionality reduction of Poisson's equation

Accordingly, Eqs. 3 and 4 are written as in Eqs. 6 and 7 for the multi-region decomposition, yielding the general formulation of the dimensionality-reduction approach.

$$\left(\frac{d^2\phi_n^x}{dx^2}\bigg|_{x=x_i}\right)L_{y,n} + \left(\int_{y_{g,n}-\frac{L_{y,n}}{2}}^{y_{g,n}+\frac{L_{y,n}}{2}}\left(\frac{\partial\phi_m^y}{\partial x}\bigg|_{x=x_i+\frac{\Delta x}{2}}\right)dy - \int_{y_{g,n}-\frac{L_{y,n}}{2}}^{y_{g,n}+\frac{L_{y,n}}{2}}\left(\frac{\partial\phi_m^y}{\partial x}\bigg|_{x=x_i-\frac{\Delta x}{2}}\right)dy\right)\frac{1}{\Delta x}$$
$$+ \left(\frac{\partial\phi_m^y}{\partial y}\bigg|_{y=y_{g,n}+\frac{L_{y,n}}{2}} + \frac{\partial\phi_n^x}{\partial y}\bigg|_{y=y_{g,n}+\frac{L_{y,n}}{2}}\right) - \left(\frac{\partial\phi_m^y}{\partial y}\bigg|_{y=y_{g,n}-\frac{L_{y,n}}{2}} + \frac{\partial\phi_n^x}{\partial y}\bigg|_{y=y_{g,n}-\frac{L_{y,n}}{2}}\right) \quad \text{(Eq. 6)}$$
$$= -\frac{1}{\epsilon_0}\int_{y_{g,n}-\frac{L_{y,n}}{2}}^{y_{g,n}+\frac{L_{y,n}}{2}} \rho(x,y)dy,$$

$$\left(\frac{d^2\phi_m^y}{dy^2}\bigg|_{y=y_j}\right)L_{x,m} + \left(\int_{x_{g,m}-\frac{L_{x,m}}{2}}^{x_{g,m}+\frac{L_{x,m}}{2}}\left(\frac{\partial\phi_n^x}{\partial y}\bigg|_{y=y_j+\frac{\Delta y}{2}}\right)dx - \int_{x_{g,m}-\frac{L_{x,m}}{2}}^{x_{g,m}+\frac{L_{x,m}}{2}}\left(\frac{\partial\phi_n^x}{\partial y}\bigg|_{y=y_j-\frac{\Delta y}{2}}\right)dx\right)\frac{1}{\Delta y}$$
$$+ \left(\frac{\partial\phi_n^x}{\partial x}\bigg|_{x=x_{g,m}+\frac{L_{x,m}}{2}} + \frac{\partial\phi_m^y}{\partial x}\bigg|_{x=x_{g,m}+\frac{L_{x,m}}{2}}\right) \quad \text{(Eq. 7)}$$
$$- \left(\frac{\partial\phi_n^x}{\partial x}\bigg|_{x=x_{g,m}-\frac{L_{x,m}}{2}} + \frac{\partial\phi_m^y}{\partial x}\bigg|_{x=x_{g,m}-\frac{L_{x,m}}{2}}\right) = -\frac{1}{\epsilon_0}\int_{x_{g,m}-\frac{L_{x,m}}{2}}^{x_{g,m}+\frac{L_{x,m}}{2}} \rho(x,y)dx.$$

In Eqs. 6 and 7, $L_{x,m}$ and $L_{y,n}$ are the horizontal and vertical extents of each region $\Omega_{n,m}$, $x_i$ is the x-coordinate of each computational node $i$ along a horizontal grid and $y_j$ is similarly the y-coordinate of each node $j$ along a vertical grid.

Having solved the coupled system of equations given by Eqs. 6 and 7 with the appropriate boundary conditions, to be described shortly, the electric field is determined within each region using $\vec{E} = -\vec{\nabla}\phi(x,y) = -\vec{\nabla}\left(\phi_n^x(x) + \phi_m^y(y)\right)$.

In Eqs. 6 and 7, it is important to note the appearance of the so-called "cross-derivative" terms, $\frac{\partial \phi_n^x}{\partial y}$ and $\frac{\partial \phi_m^y}{\partial x}$, that take into account the variations in $\phi_n^x$ and $\phi_m^y$ when transiting from one region to another along the y- and x-direction, respectively.

Moreover, recalling that, inside each region $\Omega_{n,m}$, $\phi_n^x$ is, by definition, constant along y, and $\phi_m^y$ is likewise constant along x, the integral terms within the second parentheses on the left-hand-side of Eqs. 6 and 7 are zero for the computational cells that reside entirely inside each region. In other words, the integrand terms $\frac{\partial \phi_m^y}{\partial x}\big|_{x=x_i+\frac{\Delta x}{2}}$ and $\frac{\partial \phi_m^y}{\partial x}\big|_{x=x_i-\frac{\Delta x}{2}}$ in Eq. 6 and, similarly, $\frac{\partial \phi_n^x}{\partial y}\big|_{y=y_j+\frac{\Delta y}{2}}$ and $\frac{\partial \phi_n^x}{\partial y}\big|_{y=y_j-\frac{\Delta y}{2}}$ in Eq. 7 are *non-zero* only when one side of a computational cell falls on a region's boundary.

The boundary conditions accompanying Eqs. 6 and 7 are of the following general form

$$\begin{cases} \phi_n^x(x_b) + \phi_m^{y0} = \frac{1}{L_{y,n}} \int_{y_{g,n}-\frac{L_{y,n}}{2}}^{y_{g,n}+\frac{L_{y,n}}{2}} \phi(x_b, y)\, dy, \\ \phi_m^y(y_b) + \phi_n^{x0} = \frac{1}{L_{x,m}} \int_{x_{g,m}-\frac{L_{x,m}}{2}}^{x_{g,m}+\frac{L_{x,m}}{2}} \phi(x, y_b)\, dx, \end{cases}$$

(Eq. 8)

in which, $x_b$ and $y_b$ are the locations of the domain boundary along the x and y-direction, respectively. In addition,

$$\phi_m^{y0} = \frac{1}{L_{y,n}} \int_{y_{g,n}-\frac{L_{y,n}}{2}}^{y_{g,n}+\frac{L_{y,n}}{2}} \phi_m^y(y) dy, \quad \text{and,} \tag{Eq. 9}$$

$$\phi_n^{x0} = \frac{1}{L_{x,m}} \int_{x_{g,m}-\frac{L_{x,m}}{2}}^{x_{g,m}+\frac{L_{x,m}}{2}} \phi_n^x(x) dx. \tag{Eq. 10}$$

The formulation presented above for the dimensionality-reduction approach provides a generalizable "quasi-multi-dimensional" approximation of Poisson's problem, which is included in the PIC scheme to yield the reduced-order PIC.

### 2.2. Numerical implementation of the reduced-order PIC scheme and its computational benefit

The reduced-order PIC scheme is currently implemented as an electrostatic, explicit code, which is named IPPL (Imperial Plasma Propulsion Laboratory) code. The IPPL code is written entirely in Julia language [12]. It, overall, follows the standard algorithm of the particle-in-cell method [13][14]. The main difference between the reduced-order PIC code and a conventional one pertains to the module solving the electric potential. This module, which we term the "reduced-dimension" Poisson solver, is explained in the following. The other difference is in the module that scatters the particle-based data (such as the electric charge) onto the grids and gathers the grid-based data (such as the electric field) onto the particles' position. The particles' information in the reduced-order PIC is scattered onto the computation grids corresponding to the dimensionality-reduction approach, i.e., in the most general case, $N$ horizontal grids along y and $M$ vertical grids along x. Similarly, the field information used to update the particles' velocity and position are gathered from these computation grids.

The reduced-dimension Poisson solver numerically solves the system of equation given by Eqs. 6 and 7 using Julia's built-in direct matrix solve algorithm based on the LU decomposition method. However, for the solver's implementation in the PIC scheme, instead of directly solving the system of coupled Eqs. 6 and 7, the terms corresponding to the gradients of the potential function $\phi_m^y$ in Eq. 6 and, similarly, the terms representing the gradients of $\phi_n^x$ in Eq. 7 are moved to the right-hand side and are, thus, calculated from the previous timestep of the simulation. In this way, the system of equation is kind of decoupled, which is computationally beneficial as the size of the matrix to be solved at each time step is reduced by half.

It is also important to mention that, before calculating the right-hand-side derivative terms specified in the preceding paragraph and solely for the purpose of these calculations, a moving average filter with a pre-fixed window size is applied to the solutions of potential functions $\phi_n^x$ and $\phi_m^y$. The analysis of the sensitivity of the reduced-order simulation results to the size the filter's window is presented in Section A.1 of the Appendix.

Following the above description of the numerical implementation of the reduced-order PIC scheme, we demonstrate quantitatively below the computational advantage that our novel PIC scheme has over traditional simulations in a 2D case. Table 1 provides a comparison between the full-2D and the quasi-2D PIC schemes in terms of two simulation parameters, the number of cells ($N_{cells}$) and the total number of macroparticles at the beginning of the simulation ($N_{p,init}$).

|  | $N_{cells}$ | $N_{PPC}$ | $N_{p,init}$ |
|---|---|---|---|
| **Full-2D** | $N_i \times N_j = 128,000$ | 150 | $1.92 \times 10^7$ |
| **Quasi-2D** (Single region) | $\max(N_i, N_j) = 500$ | 150 | $7.5 \times 10^4$ |
| **Quasi-2D** (N × M regions) (N = 20, M = 40) | $\max(N \times N_i, M \times N_j) = 20,000$ | 150 | $3 \times 10^6$ |

Table 1: Comparison between a full-2D and quasi-2D PIC simulation in terms of the grid size ($N_{cells}$) and the total number of macroparticles at the simulation start ($N_{p,init}$); $N_{PPC}$ is the number of macroparticles per cell.

For the comparison in Table 1, we have considered a domain with $N_i = 500$ cells along the x-direction and $N_j = 256$ cells along the y-direction. These are the same number of computational cells used for the full-2D axial-azimuthal benchmark of Ref. [15], with which we will later compare the results of our reduced-order quasi-2D simulations. Looking at the last column of Table 1, the single-region quasi-2D PIC results in 256 times reduction in the computational cost. Moreover, decomposing the domain into multiple regions, e.g., $N = 20, M = 40$ which as will be seen in Section 5 practically provides the full-2D results in all respects, still translates into a computational gain by a factor of about 6.

### Section 3: Verification of the mathematical consistency of the dimensionality-reduction formulation

The discussions in this section are aimed at verifying the mathematical consistency and generalizability of the dimensionality-reduction approach itself. In this regard, we start by presenting the proof that the formulation developed for the dimensionality reduction, discussed in the Section 2.1, mathematically converges to the 2D Poisson's equation in the limit where the number of regions is the same as the number of 2D computational cells, i.e., $N = N_j$ and $M = N_i$, with $N_j$ being the cell number along the y-direction, and $N_i$ being the cell number along the x-direction.

Next, we show, in Section 3.2, the standalone verifications of the reduced-dimension Poisson solver. We intend to demonstrate that low-number-of-region approximations of a 2D Poisson problem are sufficiently representative of the problem's full-2D solution and that, by increasing the number of regions, the approximate solutions from the reduced-dimension Poisson solver converge to the 2D solution.

### 3.1. Proof of convergence to 2D Poisson's equation

We start by recalling the 2D Poisson's equation

$$\nabla^2 \phi(x,y) = \frac{\partial^2 \phi(x,y)}{\partial x^2} + \frac{\partial^2 \phi(x,y)}{\partial y^2} = -\frac{\rho(x,y)}{\epsilon_0}. \quad \text{(Eq. 11)}$$

As pointed out in Section 2.1, the dimensionality-reduction technique states that the 2D potential distribution *within each generic region* $\Omega_{n,m}$ can be expressed in terms of a linear superimposition of two potential functions $\phi_n^x$ and $\phi_m^y$, thus, $\phi(x,y) = \phi_n^x + \phi_m^y$. Accordingly, the 2D Poisson's equation can be rewritten as follows in terms of the potential functions $\phi_n^x$ and $\phi_m^y$

$$\nabla^2 \phi(x,y) = \nabla^2 \phi_n^x(x) + \nabla^2 \phi_m^y(y) = \left(\frac{\partial^2 \phi_n^x}{\partial x^2} + \frac{\partial^2 \phi_n^x}{\partial y^2}\right) + \left(\frac{\partial^2 \phi_m^y}{\partial x^2} + \frac{\partial^2 \phi_m^y}{\partial y^2}\right). \quad \text{(Eq. 12)}$$

Rearranging the terms inside the parentheses in Eq. 12, we have

$$\left(\frac{\partial^2 \phi_n^x}{\partial x^2} + \frac{\partial^2 \phi_m^y}{\partial x^2}\right) + \left(\frac{\partial^2 \phi_n^x}{\partial y^2} + \frac{\partial^2 \phi_m^y}{\partial y^2}\right) = -\frac{\rho(x,y)}{\epsilon_0}. \quad \text{(Eq. 13)}$$

Now, we refer to Figure 3, which depicts the schematics of two computational elements associated with the dimensionality-reduction approach. The graph on the left-hand side is the enlarged view of a computational element along the x-direction within a region $\Omega_{n,m}$, as seen before in Figure 2, with the width $\Delta x$ and the length $L_{y,n}$. Similarly, the graph on the right-hand side of Figure 3 is the enlarged view of a computational element along the y-direction with the width $\Delta y$ and the length $L_{x,m}$. $x$ and $y$ denote the positions of a specific computational node along the x- and y-direction.

In Figure 3, the terms $E_N$, $E_S$, $E_E$ and $E_W$ represent, respectively, the electric field components along the North, South, East, and West sides of a computational element. The definition of each electric field component, shown in Figure 3, is consistent with the definition presented in Section 2.1, i.e., $\vec{E} = -\vec{\nabla}\left(\phi_n^x(x) + \phi_m^y(y)\right)$. The colored differential terms are the "cross-derivative" terms, with the blue ones being the terms that are always non-zero in Eqs. 6 and 7, whereas the terms in red are only non-zero when one side of the computational element falls on a region's boundary.

At this point, we recall the integral form of the Gauss' law, as given by Eq. 14,

$$\int_{\delta\Omega_{n,m}} \vec{E} \cdot d\vec{A} = \frac{Q}{\epsilon_0}. \tag{Eq. 14}$$

in which $\vec{E}$ is the electric field vector, $\vec{A}$ is the vector representing the surface area of a computational element, and $Q$ is the total charge enclosed within the computational volume. $\delta\Omega_{n,m}$ denotes the boundary of the region $\Omega_{n,m}$.

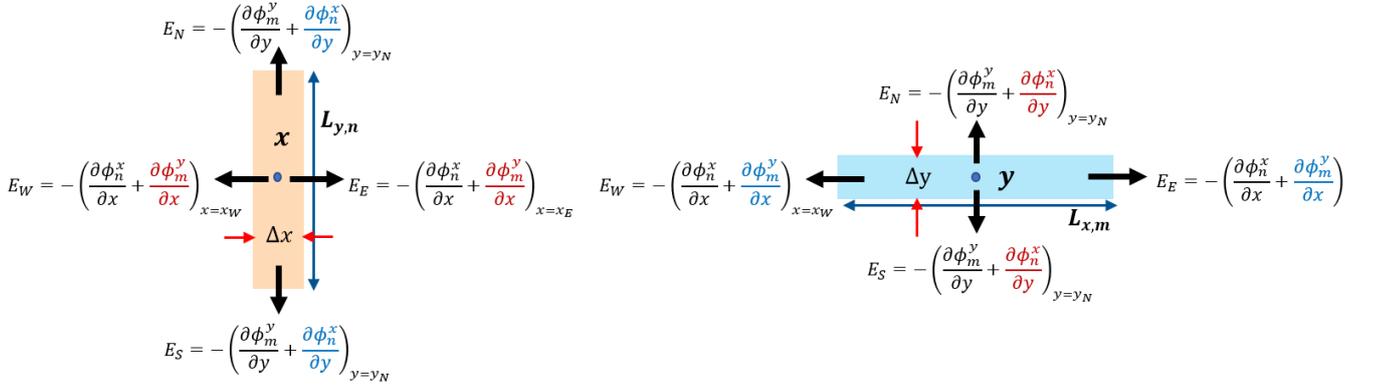

Figure 3: Schematics of the horizontal and vertical computational elements associated with the dimensionality-reduction approach; (left) the vertical element for solving potential function $\phi_n^x$ along the x-direction, (right) the horizontal element for solving potential function $\phi_m^y$ along the y-direction.

Noting the definitions provided in the preceding paragraphs and illustrated in Figure 3, the Gauss's law can be written as in Eqs. 15 and 16 for each 2D horizontal and vertical computational element, respectively,

$$(E_E - E_W)L_{y,n} + (E_N - E_S)\Delta x = \frac{Q^x(x)}{\epsilon_0} = \frac{1}{\epsilon_0}\rho^x(x)\Delta x L_{y,n}, \tag{Eq. 15}$$

$$(E_E - E_W)\Delta y + (E_N - E_S)L_{x,m} = \frac{Q^y(y)}{\epsilon_0} = \frac{1}{\epsilon_0}\rho^y(y)\Delta y L_{x,m}. \tag{Eq. 16}$$

In Eqs. 15 and 16, $Q^x$ and $Q^y$ are the total charge averaged over the y- and x-direction, respectively, and $\rho^x$ and $\rho^y$ are the corresponding charge densities.

As we increase the number of regions, the lengths of the regions, $L_{y,n}$ and $L_{x,m}$ decrease. In the limit of having as many regions as there are computational cells, the length $L_{y,n}$ approaches $\Delta y$ and the length $L_{x,m}$ approaches $\Delta x$. In this situation, the two computational elements shown in Figure 3 effectively collapse on each other for each computational node with positions $x$ and $y$ within the 2D plane. Consequently, $\rho^x$ and $\rho^y$ become equal to $\rho(x,y)$, which is the charge density at the computational node with coordinates $(x,y)$. Accordingly, in the limit, the Gauss' law for any computational element is the summation of Eqs. 15 and 16, which can be written as below by substituting the definition of each electric field component from Figure 3,

$$-2\left(\left(\frac{\partial \phi_n^x}{\partial x}+\frac{\partial \phi_m^y}{\partial x}\right)_{x=x_E}-\left(\frac{\partial \phi_n^x}{\partial x}+\frac{\partial \phi_m^y}{\partial x}\right)_{x=x_W}\right)\Delta y$$
$$-2\left(\left(\frac{\partial \phi_m^y}{\partial y}+\frac{\partial \phi_n^x}{\partial y}\right)_{y=y_N}-\left(\frac{\partial \phi_m^y}{\partial y}+\frac{\partial \phi_n^x}{\partial y}\right)_{y=y_N}\right)\Delta x = 2\frac{\rho(x,y)}{\epsilon_0}\Delta x\,\Delta y.$$ (Eq. 17)

Dividing both sides of Eq. 17 by 2 and by $\Delta x\,\Delta y$, it is seen in Eq. 18 that we recover the 2D Poisson's equation (Eq. 13), thus, proving the mathematical consistency of the dimensionality-reduction formulation

$$\left(\frac{\partial^2 \phi_n^x}{\partial x^2}+\frac{\partial^2 \phi_m^y}{\partial x^2}\right)+\left(\frac{\partial^2 \phi_n^x}{\partial y^2}+\frac{\partial^2 \phi_m^y}{\partial y^2}\right)=-\frac{\rho(x,y)}{\epsilon_0}.$$ (Eq. 18)

### 3.2. Verification of the reduced-dimension Poisson solver for general 2D problems

Three canonical two-dimensional Poisson problems are chosen. The definitions of these three problems together with their corresponding boundary conditions are presented in Table 2. In all cases, the domain is a Cartesian ($x-y$) plane of unit length along the x and y directions. For simplicity, the permittivity of free space ($\epsilon_0$) is assumed to be one for all problems. The number of computational cells for all cases are the same and equal to 200, i.e., $N_i = N_j = 200$.

| Case No. | Definition | Boundary Conditions | Reference |
|---|---|---|---|
| 1 | $\phi_{xx}+\phi_{yy}=x^2+y^2$ | $\phi(x,0)=0$ ; $\phi(x,L_y)=\frac{1}{2}x^2$ <br> $\phi(0,y)=\sin(\pi y)$ <br> $\phi(L_x,y)=\exp(\pi L_x)\sin(\pi y)+\frac{1}{2}y^2$ | [16] |
| 2 | $\phi_{xx}+\phi_{yy}=y\sin(5\pi x)+\frac{1}{0.02}\exp\left(-(x-0.5)^2+(y-0.5)^2\right)$ | $\phi(x,0)=\phi(x,L_y)=0$ <br> Periodic boundary along $y$ | [17] |
| 3 | $\phi_{xx}+\phi_{yy}=\sin(10\pi y)\sin(10\pi x)$ | $\phi(x,0)=\phi(x,L_y)=0$ <br> $\phi(0,y)=\phi(L_x,y)=0$ | --- |

Table 2: Definition of the 2D Poisson problems selected to verify the dimensionality-reduction formulation

Referring to Table 2, it is worth pointing out the reasons for the selection of these three specific Poisson problems. Regarding Case 1, the problem has a rather complex analytical solution $\phi(x,y)=\exp(\pi x)\sin(\pi y)+\frac{1}{2}(xy)^2$. As a result, the ability of the reduced-dimension Poisson solver to provide accurate numerical approximations of this solution proves the versatility of the approach. Moreover, the importance of Case 2 lies in its periodic boundary condition along the y direction. In this respect, as this boundary condition is a typical one applied along the azimuth in Hall thruster simulations that involve this coordinate, it was deemed necessary to demonstrate the applicability of the reduced-dimension solver to the problems with periodic boundary condition as well. Finally, Case 3 represent a problem with an oscillatory charge distribution of relatively small wavelength. This is a common situation occurring in a Hall thruster discharge, especially along the azimuthal coordinate where small-wavelength instability-induced fluctuations in charge density develop. This test case verifies the ability of the approach to accurately resolve the potential distribution in the presence of small-scale density gradients.

Figure 4 and Figure 5 show, respectively, the approximate solutions of the Poisson problems in Case 1 and Case 2 from the reduced-dimension Poisson solver for various number of regions $N$ and $M$. Plots (a) to (c) in both figures correspond to the evolution of the approximate solution when $N$ and $M$ are increased simultaneously from 5 to 50 regions. The plots represent the reconstructed 2D potential distribution from the solutions of potential functions $\phi_n^x$ and $\phi_m^y$. Plot (d) in Figure 4 illustrates the analytical solution for Case 1 ($\phi(x,y)=\exp(\pi x)\sin(\pi y)+\frac{1}{2}(xy)^2$) whereas, in Figure 5, it depicts the full-2D solution of Case 2 Poisson problem obtained using Julia's direct Poisson solver function.

Looking at the plots in Figure 4, it is observed that, despite the complex nature of the analytical solution, the reduced-dimension Poisson solver is able to capture all of the solution's main features (e.g., the symmetry in the y-direction and the convex curvature along x) even at a low number of regions of $M=N=5$ (Figure 4(a)). The same is true for Case 2 (Figure 5) where the approximate solution obtained using only 5 horizontal and vertical regions is seen to quite reasonably resemble the full-2D solution in Figure 5(d).

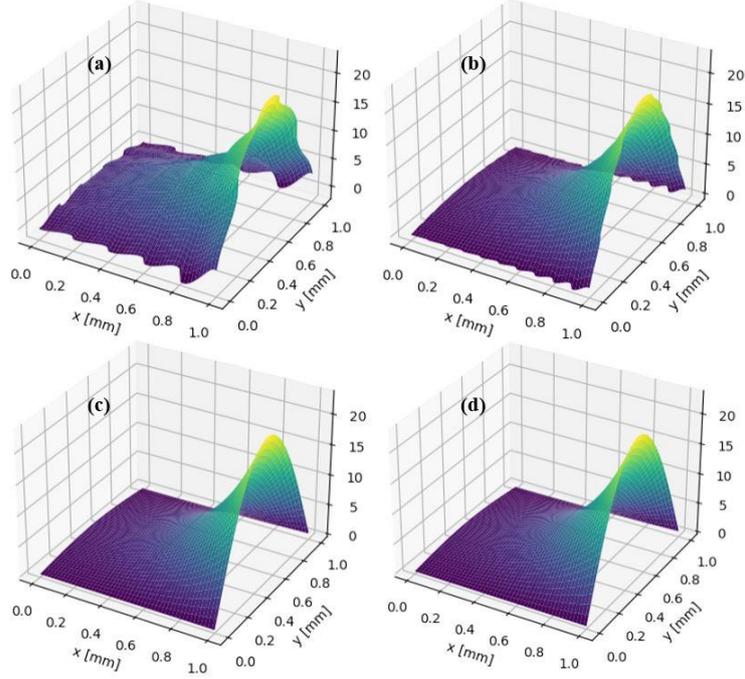

Figure 4: Numerical approximations of the solution of Poisson problem in Case 1 from the reduced-dimension solver for various number of regions, (a) $M = N = 5$, (b) $M = N = 20$, (c) $M = N = 50$. Plot (d) depicts the analytical solution $\phi(x, y) = \exp(\pi x) \sin(\pi y) + \frac{1}{2}(xy)^2$.

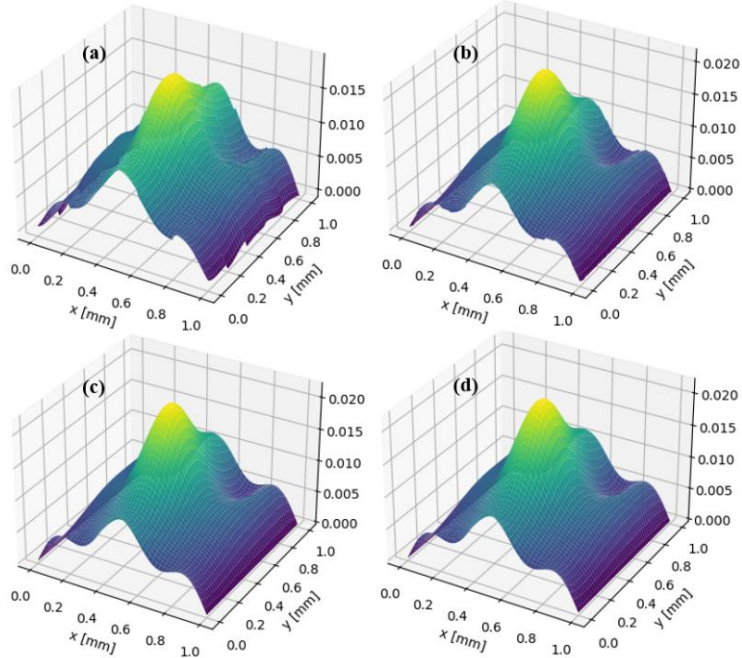

Figure 5: Numerical approximations of the solution of Poisson problem in Case 2 from the reduced-dimension solver for various number of regions, (a) $M = N = 5$, (b) $M = N = 20$, (c) $M = N = 50$. Plot (d) depicts the solution obtained using Julia's full-2D direct Poisson solver function based on the Gaussian elimination algorithm.

Furthermore, in both Cases 1 and 2, as the number of regions $N$ and $M$ are increased, the approximate solution is seen to evolve toward the "ground-truth" solution such that, at $M = N = 50$, the reconstructed 2D solution from the reduced-dimension Poisson solver has essentially converged to the "true" one. The same converging behavior of the solution is observed for the Poisson problem in Case 3 (Figure 6). Indeed, from the right-hand-side plot of Figure 6, which shows the distribution of error, as defined by Eq. 19, between the approximate solution at $M =$ $N = 100$ (Figure 6(left)) and the full-2D solution, the maximum error is found to be $\pm 1.2\%$, which is in the same order of the error corresponding to the full-2D solver itself.

$$err = \frac{1}{\max(\phi_{2D}(i,j))}(\phi_{2D}(i,j) - \phi_{Q2D}(i,j)) \qquad \text{(Eq. 19)}$$

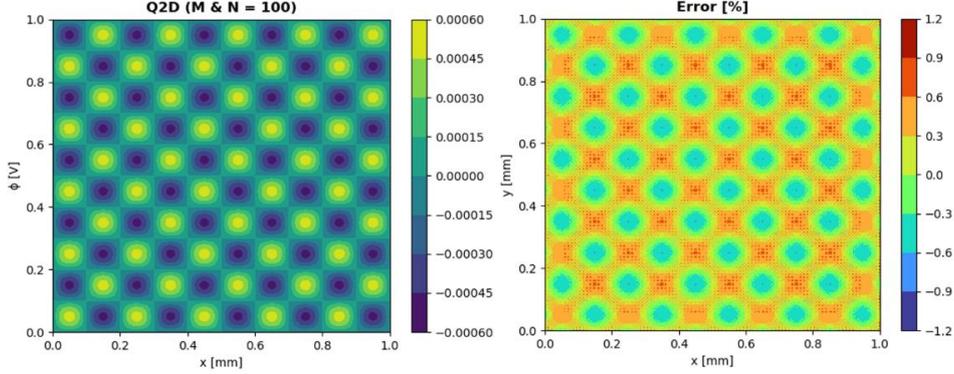

Figure 6: (left) Solution of the Poisson problem in Case 3 from the reduced-dimension Poisson solver in the limit of $M = N = 100$; (right) distribution of the error between the approximate solution and the one from the full-2D solver.

The convergence characteristic of the reduced-dimension Poisson solver and the decrease in error between the approximate and ground-truth solutions when increasing the number of regions can be better understood by looking at the plots of Figure 7, which present the evolution of the L2 error (Eq. 20) for different Poisson problems.

$$err_{L2} = \sqrt{\sum_{j=1}^{N_j}\sum_{i=1}^{N_i}\frac{\left(\phi_{2D}(i,j) - \phi_{Q2D}(i,j)\right)^2}{N_j N_i}} \qquad \text{(Eq. 20)}$$

It is observed that, in Case 1 and 2, the L2 error falls rapidly when increasing the number of regions, approaching error values in the same order of a full-2D Poisson solver. For Case 3 (Figure 7(c)), even though the error falls below $10^{-5}\%$ from about $M = N = 62$, the decrease in L2 error is rather linear in contrast to the other two cases. However, this is due to the extreme nature of charge density gradient in Case 3, which implies that higher-number-of-region approximations are necessary to provide solutions that are closely representative of the full-2D.

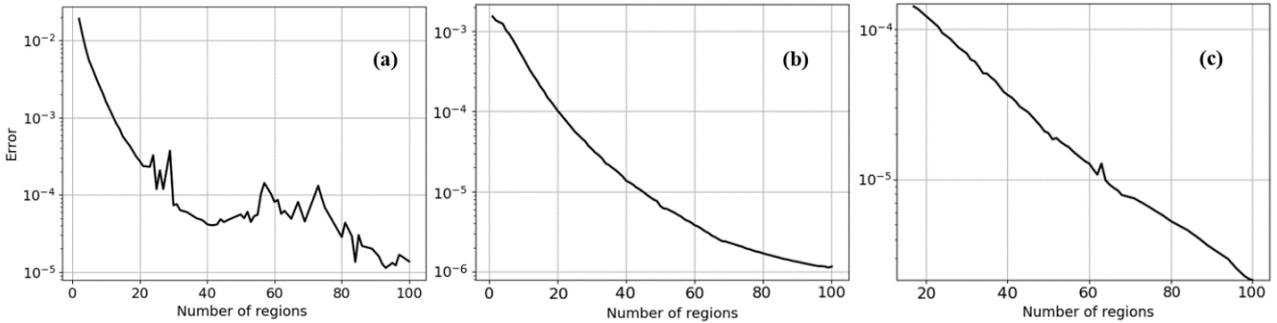

Figure 7: Variation of the L2 error (Eq. 20) between the analytical/full-2D solution and the approximate solution from the reduced-dimension Poisson solver when increasing $N$ and $M$; (a) Case 1, (b) Case 2, (c) Case 3. The y-axis in all plots is in logarithmic scale.

As the final point, the reason why the number of regions in the test cases of this section, which all corresponded to a 2D $200 \times 200$ grid, were not increased beyond 100 is that our numerical implementation of the dimensionality-reduction formulation requires that at least two nodes fall within each region.

### Section 4: Verification of the reduced-order PIC scheme in quasi-2D axial-azimuthal Hall thruster simulations

From this section forward, we focus on presenting the verification results of the reduced-order PIC scheme in an axial-azimuthal Hall thruster configuration. To this end, we have adopted the well-established 2D axial-azimuthal simulation case of Ref. [15]. The simulation setup and conditions are explained in Section 4.1.

Afterwards, we discuss the results from our quasi-2D PIC simulations in the low number-of-region limit. The aim is to demonstrate that, in this limit, the quasi-2D simulations in various simulation conditions can reproduce with notable accuracy the results from the reference full-2D simulations.

### 4.1. Overview of the simulation setup and conditions

The setup of the quasi-2D axial-azimuthal simulations is similar to that described in Refs. [15] and [18]. The simulation domain is a 2D Cartesian $(x - y)$ plane, with $x$ along the axial direction and $y$ representing the azimuthal coordinate. The adopted computational and physical parameters, including the domain's extents, cell size, time step, total simulation time and the initial conditions, are also the same as those for the full-2D reference case [15][18], which are summarized in Table 3 as well.

| Parameter | Value [unit] |
| --- | --- |
| Computational Parameters | |
| Time step ($\Delta t$) | $5 \times 10^{-12}$ [s] |
| Total simulation time ($t_{final}$) | $20 \times 10^{-6}$ [s] |
| Axial domain length ($L_x$) | 2.5 [cm] |
| Azimuthal domain length ($L_y$) | 1.28 [cm] |
| Cell size ($\Delta x = \Delta y$) | $5 \times 10^{-3}$ [cm] |
| Initial number of macroparticles per cells for axial grid ($N_{ppc_x}$) | 150 |
| Initial number of macroparticles per cells for azimuthal grid ($N_{ppc_y}$) | 300 |
| Physical Parameters | |
| Initial plasma density ($n_{p,init}$) | $5 \times 10^{16}$ [m$^{-3}$] |
| Electron injection temperature ($T_e$) | 10 [eV] |
| Ion injection temperature ($T_i$) | 0.5 [eV] |
| Discharge Voltage ($V_d$) | 200 [V] |

Table 3: Summary of the main numerical and physical parameters used for the quasi-2D axial-azimuthal simulations

The only difference between the setup of the quasi-2D and full-2D reference simulations is related to the electrons' cathode boundary condition. In this regard, whereas, in the 2D simulations, the current-equality boundary condition (see Section 6.1 for definition) has been assumed for injecting electrons from the cathode side of the domain, we applied the quasi-neutrality condition (Section 6.1) on the cathode plane for all simulations presented in this article. The reason for this choice is that we have found the reduced-order PIC scheme to be sensitive to the excessive charge imbalance that the current-equality cathode condition can introduce in the simulation during the initial transient phase [18]. We have presented, in Section 6.1, a detailed assessment of the influence of the cathode boundary condition on the quasi-2D simulation results.

The plasma potential in the quasi-2D simulations is solved using the reduced-dimension Poisson solver, explained in Section 2.2. The associated boundary conditions, whose general form were given by Eq. 8 in Section 2.1, are specified for the PIC simulations as follows: along the azimuthal computation grid(s), a periodic boundary condition is considered, which is represented as a zero-volt reference potential applied to both ends, i.e., $\phi_m^y(y_b) = 0$, where $y_b$ denotes the domain's boundary locations along the azimuth. This implementation of the periodic boundary condition is in line with the 2D reference simulation case [15][18].

Along the axial computation grid(s), Dirichlet boundary conditions of 200 V at the anode side ($x_a = 0$) and 0 V at the cathode side of the domain ($x_c = 2.5\ cm$) are implemented using the relations below

$$\begin{cases} \phi_n^x(x_a) + \dfrac{1}{L_{y,n}} \displaystyle\int_{y_{g,n}-\frac{L_{y,n}}{2}}^{y_{g,n}+\frac{L_{y,n}}{2}} \phi_m^y(y) dy = 200, \\ \phi_n^x(x_c) + \dfrac{1}{L_{y,n}} \displaystyle\int_{y_{g,n}-\frac{L_{y,n}}{2}}^{y_{g,n}+\frac{L_{y,n}}{2}} \phi_m^y(y) dy = 0. \end{cases} \quad \text{(Eq. 21)}$$

The quasi-2D simulations in the low number-of-region limit are performed across a range of simulation conditions. In this regard, it is pointed out that, similar to the reference 2D case [15][18], quasi-2D simulations feature a Gaussian magnetic field profile and a cosine-shaped, time-invariant imposed ionization source. The nominal selection of the magnetic field profile and the ionization source, which is referred to in Ref. [18] as the "benchmark" conditions, are shown in the plots (a) and (c) of Figure 8. The peak magnetic field intensity in the benchmark condition is $100\ G$, whereas the benchmark's maximum current density corresponding to the peak of the ionization source is $400\ A/m^2$ [15]. The various simulation conditions, for which the quasi-2D results are presented and compared in the following sections against the full-2D results, are obtained by scaling the profiles of the ionization rate and the magnetic field intensity, as illustrated in plots (b) and (d) of Figure 8. In particular, the simulation conditions correspond to two sets of cases: the first set includes different current densities (i.e., $J = $ 50, 100, 200 and 400 A/m$^2$) with the magnetic field peak intensity being fixed at $100\ G$. The second set is related to various magnetic field peak intensities (i.e., $B = $ 50, 100, 150 and 200 G), with the maximum current density maintained at $400\ A/m^2$.

The "low-order" quasi-2D simulations are performed with different number of regions along the axial direction, namely, $M = 1$ (single-region), $M = 2$ (double-region) and $M = 3$ (triple-region). A single region along the azimuthal direction (i.e., $N = 1$) is assumed in all cases. In the double-region simulations, the domain is divided equally into two regions along the axial direction so that the boundary of the regions falls at $x = 1.25\ cm$. In the triple-region simulations, the boundary between the first and the second region is chosen to be at the axial position of the peak magnetic field, which coincides with the exit plane of the thruster's channel in the adopted simulation domain.

Comparison against the full-2D results in terms of the time-averaged profiles of the macroscopic plasma properties are carried out for both sets of simulation conditions, as described above. These results are presented in Section 4.2. Next, we have focused on the results from the quasi-2D simulations with varying current densities and have investigated more closely the capability of the reduced-order scheme to capture the relevant interplay between the axial and azimuthal physical processes. In this respect, in Section 4.3, the characteristics of the resolved azimuthal instabilities are discussed, and the evidence that the quasi-2D double- and triple-region simulations can resolve the downstream convection of the azimuthal waves are presented. In addition, in Section 4.4, the contribution of various terms in the electrons' azimuthal momentum equation from the quasi-2D simulations are shown.

Before proceeding to the results, it is important to clarify the reason behind the fact that the resolved azimuthal physics and the interactions are only discussed for the simulation set corresponding to different current densities. This has been due to the fact that the theoretical dispersion relation of the azimuthal waves captured in our simulations [19] provides a clear relation between the waves' characteristics and the plasma density. As a result, it was deemed more relevant to evaluate whether the quasi-2D simulations can accurately resolve this theoretical dependency when varying the maximum current density values, which is equivalent to varying the plasma density.

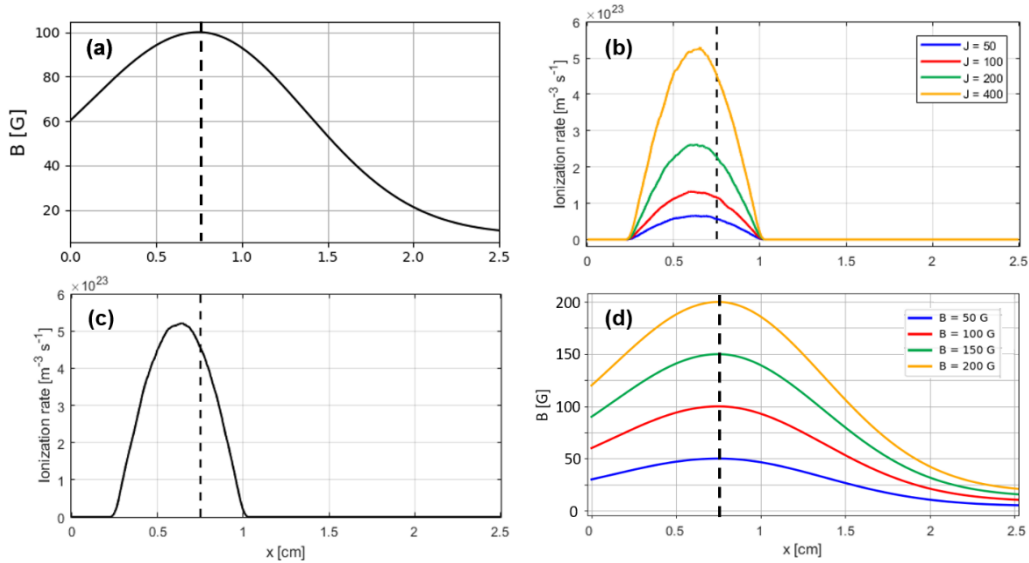

Figure 8: Axial profiles of (a) magnetic field and (b) ionization rates for the first set of quasi-2D simulations with varying current densities; Distributions of (c) the ionization source and (d) magnetic field intensities for the second set of quasi-2D simulations with varying magnetic field. The black dashed lines show the location of the peak magnetic field intensity (or the channel exit plane)

## 4.2. Prediction of the axial plasma properties' profiles

In this section, we present the predictions of the quasi-2D simulations in the low number-of-region limit in terms of the distribution of time-averaged plasma properties, ion number density ($n_i$), electron temperature ($T_e$), electric field ($E_x$), and the axial plasma potential ($\phi_1^x$) in which the subscript 1 refers to the use of only one region along the azimuthal direction and, thus, only a single axial computation grid, in the low number-of-region simulations.

The axial plasma distributions presented in the following subsections are averaged over the last $4\ \mu s$ of the simulations' time, i.e., from time $16\ \mu s$ to $20\ \mu s$.

### 4.2.1. Various current densities

Figure 9 and Figure 10 show the comparison between the axial plasma properties' profiles from our quasi-2D simulations and the 2D results of Ref. [18] for various current density values. In the ion number density plots of Figure 9, the reference 2D profiles are shown at dotted lines. In Figure 10, however, the 2D profiles of the axial electric field and the azimuthally averaged plasma potential for different current densities are shown separately in plots (g) and (h).

Referring to plots (a) and (b) in Figure 9 and in Figure 10, it is first observed that the single-region quasi-2D simulations provide a very consistent approximation of the results from the full-2D simulations. Nevertheless, the double-region and triple-region simulations demonstrate an improvement in the prediction of the axial distribution of plasma properties. In this respect, looking at plots (a), (c) and (e) in Figure 9, it is seen that the trend of ion density profiles predicted by the triple-region simulations resemble very closely the reference 2D profiles, with only a slight difference observed near the peak of the density.

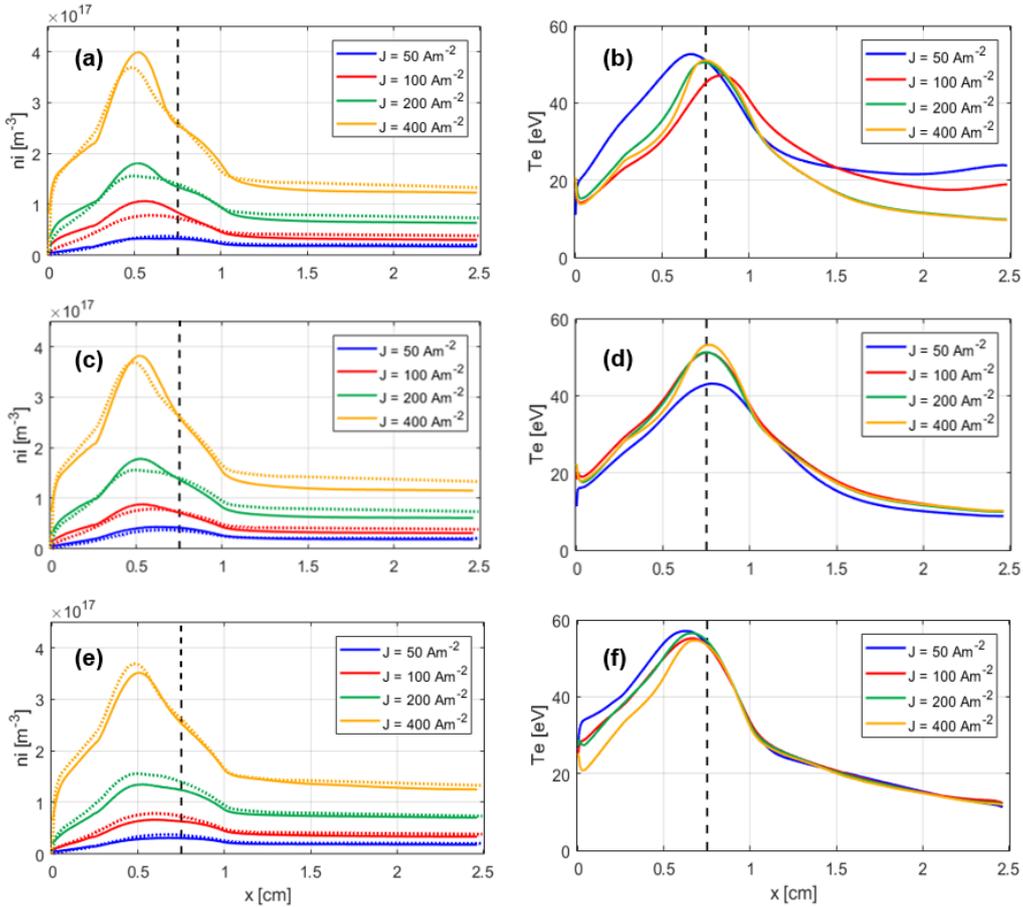

Figure 9: Time-averaged axial profiles of the ion number density (left column) and the electron temperature (right column) for various current densities from the quasi-2D simulations; plots (a) and (b) are from single-region simlation, (c) and (d) from the double-region, and (e) and (f) from the triple-region simulations. The dotted lines in the ion number density plots represent the profiles from the 2D simulations reported in Ref. [18].

Moreover, referring to the electron temperature plots on the right-hand-side of Figure 9, it is worth pointing out that, in the quasi-2D double- and triple-region simulations, the electron temperatures near the right boundary of the domain remain close to the temperature of 10 eV at which the electrons are injected from the cathode. This result is consistent with what has been observed in the 2D reference simulations [15][18]. Nonetheless, this feature had not been observed in our previous work [10] where the preliminary "pseudo-2D" simulations showed temperatures in the near-plume region that were higher than the electrons' injection temperature from the cathode. This difference is now believed to be mostly caused by the adoption of the current-equality cathode boundary condition in the pseudo-2D simulations [10]. Noting this point, and as mentioned earlier in Section 1, the sensitivity of the reduced-order quasi-2D simulations to the model used for the electrons' injection from the cathode boundary is further discussed in Section 6.1.

Regarding the distributions of the axial electric field and plasma potential in Figure 10, it is observed that, overall, the quasi-2D results are in good agreement in all cases with the 2D simulations' predictions. However, it is again noticed that, across various current densities, the triple-region simulations provide a relatively more accurate approximation of the 2D profiles, particularly concerning the value and axial location of the peak electric field intensity.

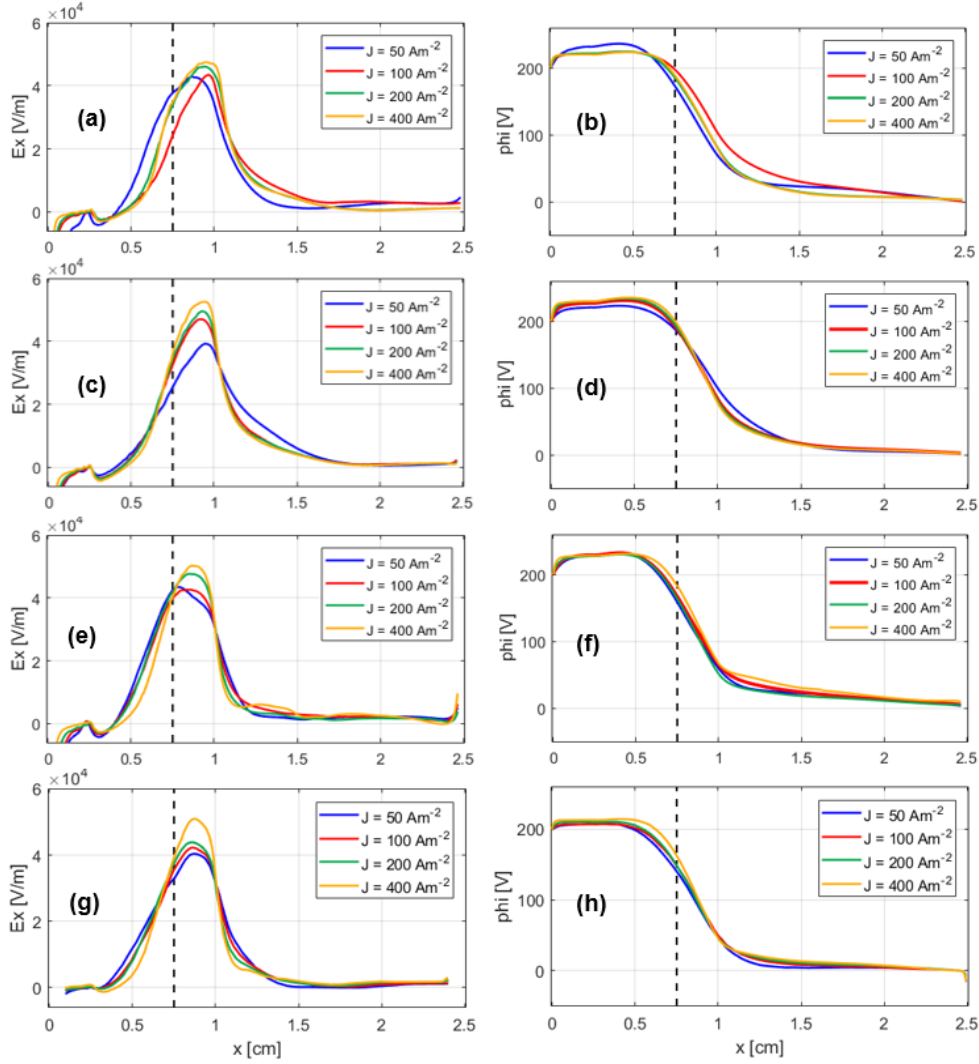

Figure 10: Time-averaged axial profiles of the axial electric field (left column) and the axial electric potential (right column) for various current densities from the quasi-2D simulations; plots (a) and (b) are from single-region simlation, (c) and (d) from the double-region, and (e) and (f) from the triple-region simulations. The corresponding full-2D reference simulation results [18] are shown in plots (g) and (h).

In Figure 11, the axial distributions of the ion Mach number from "low-order" quasi-2D simulations are presented, which indicate that, as reported for the full-2D case [15][18], ion sonic point occurs near the peak of the magnetic field for all current density values. According to the 2D simulations in Ref. [18], the ion sonic point represents the axial location at which the characteristics of the azimuthal instabilities show a "step-wise" transition. Hence, having the boundary between two adjacent regions to coincide with this location can allow the quasi-2D simulation to capture this transition in the waves' characteristics about the ion sonic point, which is, in turn, expected to improve the agreement of the quasi-2D predictions with the full-2D simulation results. In this regard, although it is observed from Figure 9 and Figure 10 that the triple-region simulations (in which the boundary between the first and the second region is at the ion sonic location) provides the most accurate predictions of the plasma profiles, it only represents a minor improvement with respect to the double-region simulations (where the boundary between the two regions was at a position further downstream the ion sonic point). This observation highlights that the predictions of the axial plasma properties are not notably sensitive to the location of regions' boundaries.

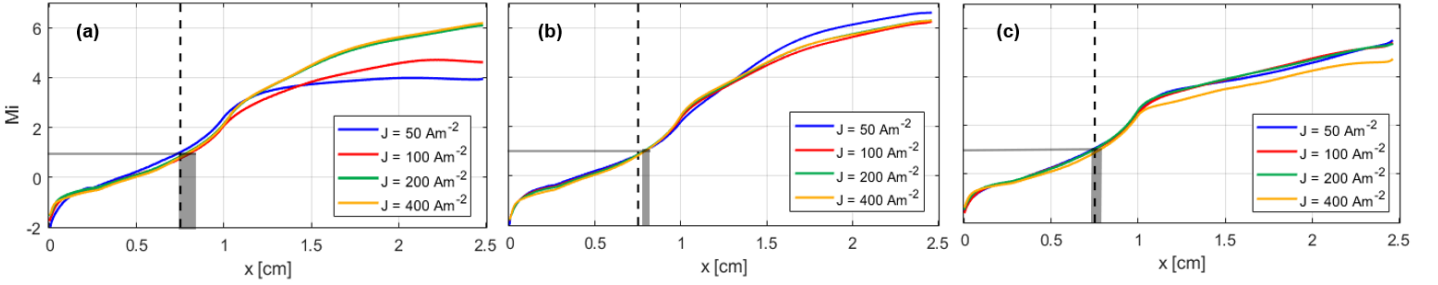

Figure 11: Axial profiles of the ion Mach number from (a) single-region, (b) double-region and (c) triple-region simulations for different current densities. The grey-shaded area represents the axial positions of the ion sonic point for various J values.

### 4.2.2. Various magnetic field intensities

Figure 12 and Figure 13 present the axial profiles of the time-averaged plasma properties for various magnetic field peak intensities obtained from the quasi-2D simulations as compared with the 2D simulation result of Ref. [18].

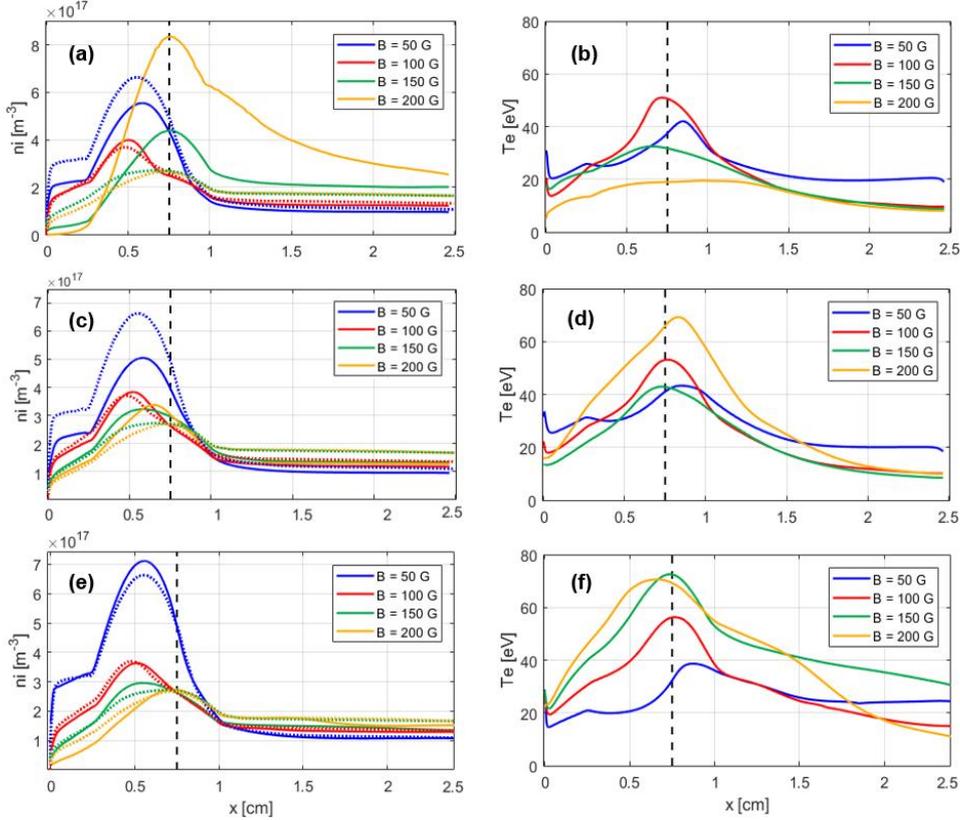

Figure 12: Time-averaged axial profiles of the ion number density (left column) and the electron temperature (right column) for various magnetic field peak intensities from the quasi-2D simulations; plots (a) and (b) are from single-region simlation, (c) and (d) from the double-region, and (e) and (f) from the triple-region simulations. The dotted lines in the ion number density plots represent the profiles from the 2D simulations reported in [18].

As it is seen in the plots (a) and (b) in these figures, the single-region simulations are unable to properly approximate the plasma properties in the 150 G and 200 G cases. In fact, in these two cases, the resolved instability-induced electron cross-field transport is not enough to allow for a sufficient electron current toward the anode. As a result, the electrons and, hence, the plasma generated in the ionization zone (defined by the extent over which the ionization source is imposed) does not fill the near-anode region. This is evident in Figure 12(a) which shows much low densities near the anode in the cases with high magnetic field peaks. Accordingly, as observed from Figure 13(b), the electric potential is not sustained at the anode voltage and drops quickly near the anode.

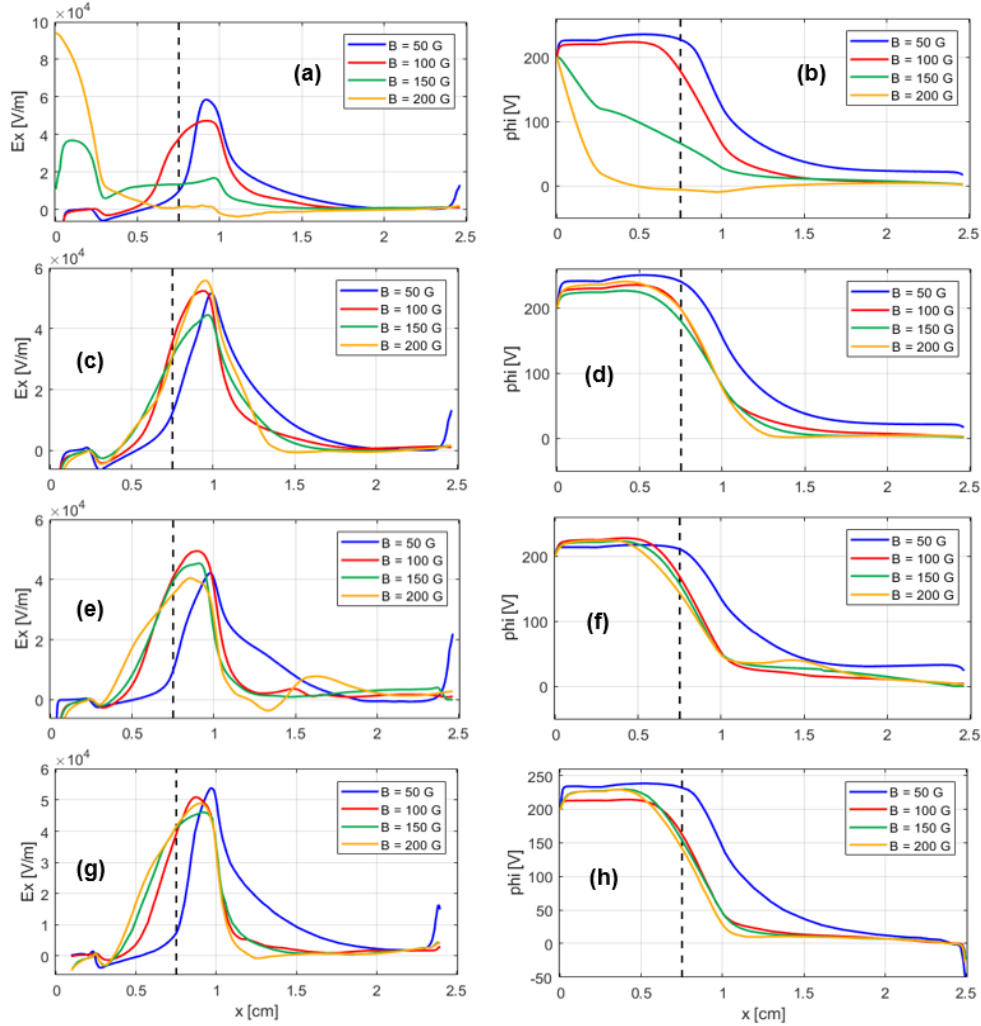

Figure 13: Time-averaged axial profiles of the axial electric field (left column) and the axial electric potential (right column) for various magnetic field peak intensities from the quasi-2D simulations; plots (a) and (b) are from single-region simlation, (c) and (d) from the double-region, and (e) and (f) from the triple-region simulations. The corresponding full-2D reference simulation results [18] are shown in plots (g) and (h).

Therefore, the high magnetic-field-peak simulation conditions represent the limiting cases for the single-region quasi-2D simulations. In any case, it is necessary to point out that this limitation of the single-region approximation might be specific to the adopted reference 2D simulation case and due to the disparity between the peak value of the imposed ionization source (400 $A/m^2$) and the magentic field peak intensities. As such, we intend to further investiagte in the future work the extent of applicability of the single-region reduced-order simulation in cases with self-consistent treatment of the ionization event.

Looking now at the plots (c) to (f) in Figure 12 and Figure 13, the double-region and triple-region simulations are, nonetheless, seen to be able to properly predict the plasma properties' profiles across the considered range of magnetic field peaks. In this regard, the plots show a reasonable consistency between the quasi-2D and full-2D simulation results. Furthermore, as it was the case for the simulations with various current densities, the triple-region simulations for different magnetic field peaks also provide results that are most comparable with the 2D profiles, especially in terms of the ion number density.

However, concerning the case with the lowest magnetic field intensity (B = 50 G), although all quasi-2D simulations provide an acceptable approximation of the distribution of plasma properties, in none of the simulations, all plasma profiles can be considered to be in total agreement with the reference 2D results. This is again conjectured to be as a consequence of the specific setup of the reference 2D simulation case, and will be assessed more closely in the future work.

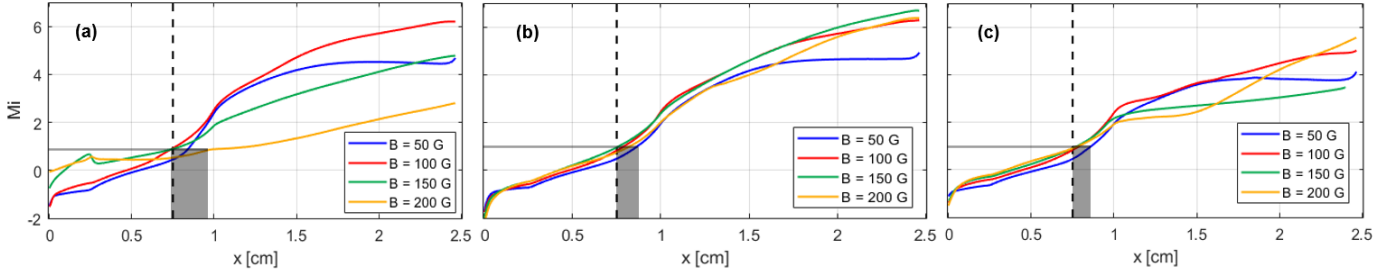

Figure 14: Axial profiles of the ion Mach number from (a) single-region, (b) double-region and (c) triple-region simulations for different magnetic field peak intensities. The grey-shaded regions represent the axial positions of the ion sonic point for various B values.

Finally, Figure 14 illustrates the axial distributions of the ion Mach number from the quasi-2D simulations for various magnetic field peak intensities. The plots indicate that, similar to what was observed for various current density cases, the ions reach the sound speed around the location of the magnetic field peak. However, increasing the number of regions $M$ along the axial direction from 1 (Figure 14(a)) to 3 (Figure 14(c)), the extent of the gray-shaded region, which represents the axial positions of the ion sonic point for different magnetic field peaks, decreases. Of course, we underline that the ion Mach number distributions for the magnetic-field-peak cases of 150 G and 200 G from the single-region simulations (Figure 14(a)) are affected by the fact that the single-region approximation in these cases could not resolve sufficient electron transport to sustain the plasma potential near the anode boundary.

**4.3. Analysis of the azimuthal waves' characteristics**

The azimuthal instabilities are demonstrated to play a significant role in regulating electrons' cross-field transport in Hall thrusters [1] and, hence, can have a remarkable influence on the distribution of plasma parameters. In the previous section, we showed that the axial profiles of the plasma properties obtained from "low-order" quasi-2D simulations are consistent with those from 2D simulations. This implies that the overall effect of the azimuthal instabilities on the axial plasma distributions is properly captured by the quasi-2D simulations. In this section, we focus on the characteristics of the instabilities that are excited in the simulations along the azimuthal direction.

One of the main instabilities exciting along the azimuthal direction in plasmas with E × B configuration, such as that in a Hall thruster, is the Electron Cyclotron Drift Instability (ECDI) [20][21]. This unstable mode features a rapid nonlinear growth, followed by a transition to ion acoustic instability which has a smaller growth rate [22][23]. The details of the initial non-linear evolution phase of the ECDI are not detectable in the present simulations and, hence, we only focus in the following on the pseudo-saturated state of the instabilities, represented by modified ion acoustic wave modes [19]. From the literature, the wavevector of the ion acoustic instability in Hall thrusters is known to be mostly along the azimuthal direction with a finite component along the axial direction as well [24].

To derive the wavenumber and frequency contents of the azimuthal fluctuations resolved in the quasi-2D simulations, we have used the MATLAB's [25] built-in 2D Fast Fourier Transform (fft2) function. The 2D FFT is applied to the last 10 $\mu s$ (i.e., from 10 to 20 $\mu s$) of the spatiotemporal signals of azimuthal electric field that are presented next. We have plotted the signals' frequency versus the azimuthal wavenumber to illustrate the numerical dispersion of the waves which is then compared against the theoretical dispersion relation of the ion acoustic instabilities in the laboratory's reference frame [26]

$$\omega_R \approx \vec{k} \cdot \vec{V}_{di} \pm \frac{k_y C_s}{\sqrt{1 + k_y^2 \lambda_D^2}},$$ (Eq. 22)

where, $\omega_R$ is the real frequency, $\vec{k}$ is the wavevector, and $k_y$ is the azimuthal wavenumber. Also, $\lambda_D$, $C_s$ and $\vec{V}_{di}$ are, respectively, the Debye length, the ion sound speed and the ions' drift velocity. The first term on the right-hand-side of Eq. 22 ($\vec{k} \cdot \vec{V}_{di}$) describes the convection of the waves with the ion flow [26][27]. As the ions' drift velocity is mostly axial, the convection is nonzero in the presence of a finite axial wavenumber ($k_x$). However, as it will be clarified in Section 5.2, in the low number-of-region quasi-2D simulations of this section, it is not possible to resolve the axial wavenumber. Therefore, in the dispersion plots shown below, we have ignored the convection term in the theoretical dispersion relation and have additionally considered the one-sided dispersion comprising of the positive frequencies only, i.e., $\omega_R \approx \frac{k_y C_s}{\sqrt{1+k_y^2 \lambda_D^2}}$.

As the final point before discussing the results related to the waves' characteristics, it is noted that, in the dispersion plots, the frequencies and wavenumbers are normalized, respectively, with respect to the average ion plasma frequency and the Debye length in a region of interest. In this respect, concerning the single-region simulations, the normalizing parameters correspond to the average over the entire axial extent of the domain.

Figure 15 shows the time evolution of the azimuthal electric field fluctuations and the corresponding dispersion plots from the single-region quasi-2D simulations for various current densities. Looking, first, at the spatiotemporal maps (left column in Figure 15), we found that the variation in waves' frequency and wavelength with the current density is in line with the results from the full-2D simulations [18]. In this respect, as the current density is increased from 50 $A/m^2$ (Figure 15, row (a)) to 400 $A/m^2$ (Figure 15, row (d)), the frequency of the waves increases whereas their wavelength decreases.

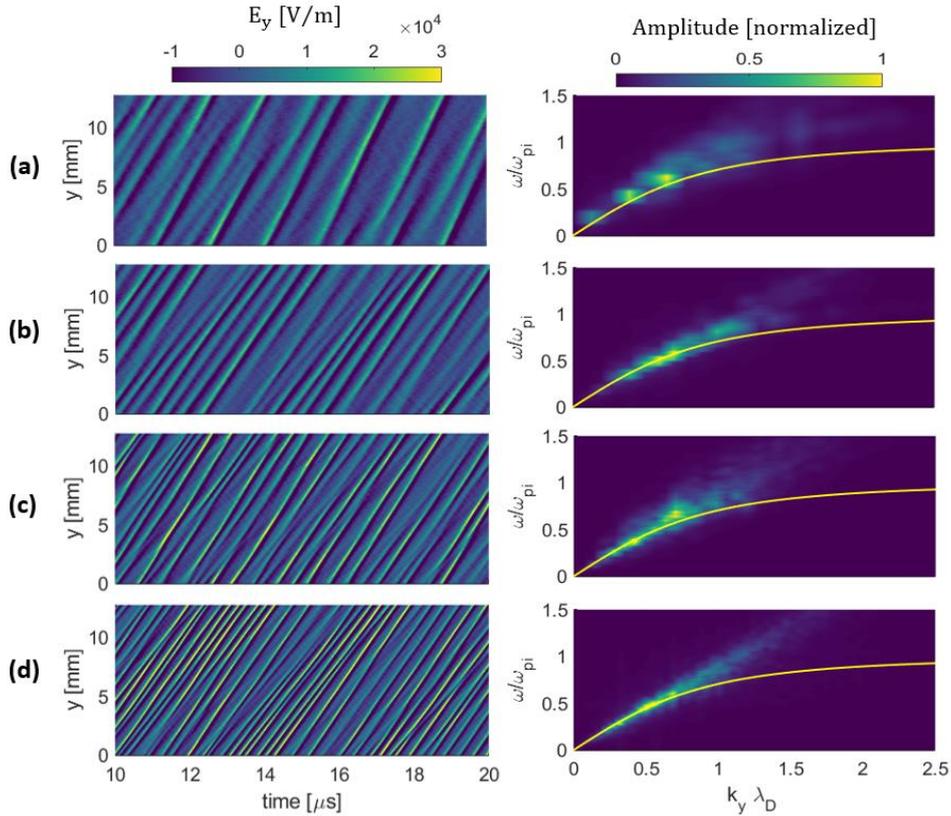

Figure 15: Spatiotemporal maps of the azimuthal electric field fluctuations (left column), and the corresponding dispersion plots (right column) from the single-region simulations for different current densities. Rows (a), (b), (c) and (d) correspond to the current densities of 50, 100, 200 and 400 $A/m^2$, respectively.

Now, referring to the dispersion plots, shown in the right-hand-side column of Figure 15, it is observed that, in all cases, the waves' dispersion from the simulations is in an almost perfect agreement with the theoretical dispersion relation of modified ion acoustic instability. Moreover, to further verify the consistency of the characteristics of the observed instabilities with those of the ion acoustic waves, we focus on the frequency and wavenumber of the most dominant mode in the dispersion plots. In this regard, it is first pointed out that, as the frequency and wavenumber are normalized with respect to the average ion plasma frequency and average Debye length in each simulation case, the loci of the most dominant wave modes remain invariant across various current density cases. Second, according to Ref. [19], the wavelength and the frequency of the resolved most dominant wave mode shall be congruous with the theoretical values corresponding with the ion acoustic instability at the maximum growth rate, i.e., $\lambda_w \approx 2\pi\sqrt{2}\lambda_D$ and $\omega_{R,max} \approx \omega_{pi}/\sqrt{3}$. The first relation in terms of the wavenumber is written as $k_y \approx 1/\sqrt{2}\lambda_D$. Accordingly, the normalized wavenumber and frequency of the most dominant mode from the theory should be $k_y\lambda_D \approx 0.71$ and $\omega_{R,max}/\omega_{pi} \approx 0.58$, which are noticed from the dispersion plots to be very similar to the values obtained from the simulations.

The above observations and analyses confirm that the single-region quasi-2D simulation provides an accurate average representation of the azimuthal waves' characteristics over various simulation conditions, which emphasizes the potentials of this simple, yet highly computationally efficient implementation of the reduced-order PIC scheme.

In Figure 16, we present the spatiotemporal maps of the azimuthal electric field fluctuations in different regions of the triple-region quasi-2D simulation for various current density conditions. Referring to the plots, it is, first, noted that, varying the current density from 50 $A/m^2$ (Figure 16, row(a)) to 400 $A/m^2$ (Figure 16, row (d)), the wavelength and the frequency of the waves are seen to change accordingly in all regions. Second, as reported in the 2D axial-azimuthal reference simulations [18], going from Region I to Region II, the characteristics of the waves show a clear change in all current-density cases. In this regard, recalling that, in the triple-region simulations, the boundary between Region I and II was set at the location of peak magnetic field intensity, which was demonstrated to coincide with the ion sonic point (Figure 11(c)), a variation in the waves' frequency and wavelength in the triple-region simulations from Region I to II, consistent with the 2D results, was expected.

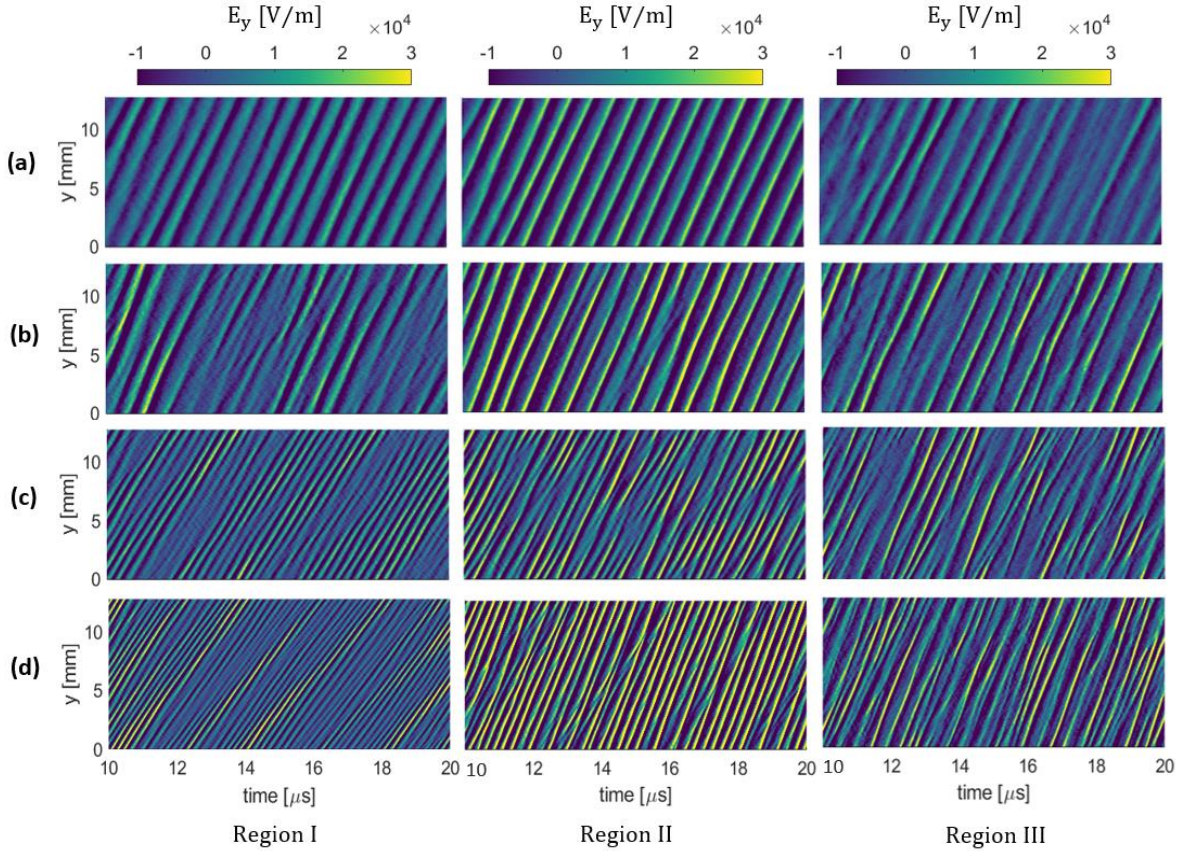

Figure 16: Spatiotemporal maps of the azimuthal electric field fluctuations in region I (left column), region II (middle column), and region III (right column) of the triple-region simulations for various current densities. Rows (a), (b), (c) and (d) correspond to the current densities of 50, 100, 200 and 400 $A/m^2$, respectively.

Referring now to Figure 17, in which the dispersion plots corresponding to the azimuthal electric field fluctuations in Figure 16 are shown, we can first confirm the statement above concerning the variation in the waves' characteristics from Region I to Region II in the triple-region simulations.

Second, an important observation to highlight here is the correlation between the resolved waves' dispersion in each region and the local theoretical dispersion relation of the ion acoustic waves (yellow lines in Figure 17) versus the theoretical dispersion relation plotted with the average plasma properties of Region I (green lines in Figure 17). In this regard, we see that the waves' dispersion map deviates increasingly from the *local* ion acoustic dispersion relation from Region I to Region II and III, whereas the dispersion characteristics remain very consistent with the theoretical dispersion relation of the ion acoustics waves in Region I. This is the evidence of the fact that, even though the "low-order" multi-region quasi-2D simulations cannot resolve the azimuthal waves' axial wavenumber, they are capable of resolving the axial convection by the ion beam of the waves excited in Region I [18], a capability that has been so far believed to be exclusive to the multi-dimensional PIC simulations.

Moreover, another interesting point to mention is that the deviation between the local ion acoustic dispersion relation and the one corresponding to Region I is essentially due to the axial component of the term $\vec{k} \cdot \vec{V}_{di}$, i.e., $k_x V_{di,x}$, in the theoretical modified ion acoustic dispersion relation (Eq. 22). Indeed, the increasing deviation from Region II to III is because of the accelerating ion beam, which implies that the average ions' axial drift velocity in Region III is higher than that in Region II.

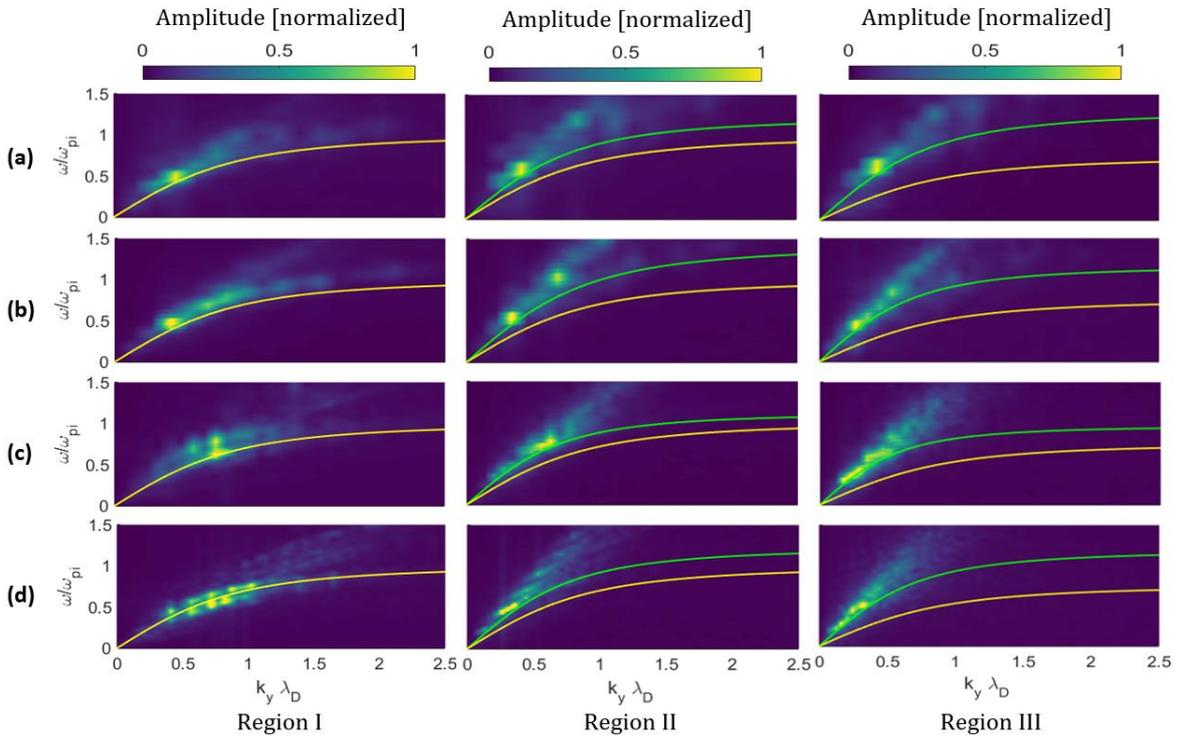

Figure 17: Dispersion plots of the azimuthal electric field oscillations in region I (left column), region II (middle column), and region III (right column) of the triple-region simulations for different current densities. Rows (a), (b), (c) and (d) correspond, respectively, to the current densities of 50, 100, 200 and 400 $A/m^2$. The yellow lines are the local dispersion relations, whereas the green lines are the dispersion relation of the ion acoustic waves in Region I.

### 4.4. Axial distribution of the force terms in the electrons' azimuthal momentum equation

To conclude the discussions on the capabilities of the "low-order" quasi-2D simulations to resolve the underlying plasma processes along the azimuthal direction and their influence on the axial plasma behavior, we refer to the plots in Figure 18. These plots show the axial profiles of the force terms in the electrons' azimuthal momentum equation, Eq. 23, from the quasi-2D simulations for various current densities. The detailed derivation of Eq. 23 can be found in Ref. [28], and the mathematical definition of each of the constituent terms in this equation is provided in Ref. [10].

$$-q n_e v_{e,x} B = \partial_t(m n_e v_{e,z}) + \partial_x(m n_e v_{e,x} v_{e,z}) + \partial_x(\Pi_{e,xz}) - q n_e E_z \qquad \text{(Eq. 23)}$$

In Eq. 23, the term on the left-hand side is referred to as the magnetic force term ($F_B$), which represents the aggregate effect of various transport mechanisms in the azimuthal momentum equation on the electrons' axial mobility (note that the electron axial drift velocity, $v_{e,x}$, appears in the relation for the magnetic force term). On the right-hand side of Eq. 23, the first term is called the temporal inertia term ($F_t$), which is not shown in the plots of Figure 18 since its contribution to transport was almost zero in all simulation cases. The second term on the right-hand side of Eq. 23 is the convective inertia ($F_I$), the third term is the viscous force term ($F_\Pi$) that captures the effect of off-diagonal terms in the pressure tensor, and the last term is the electric force term ($F_E$). As elaborated on in Ref. [10], the electric force term corresponds in our simulations to the effect of the azimuthal instabilities on the electrons' axial mobility.

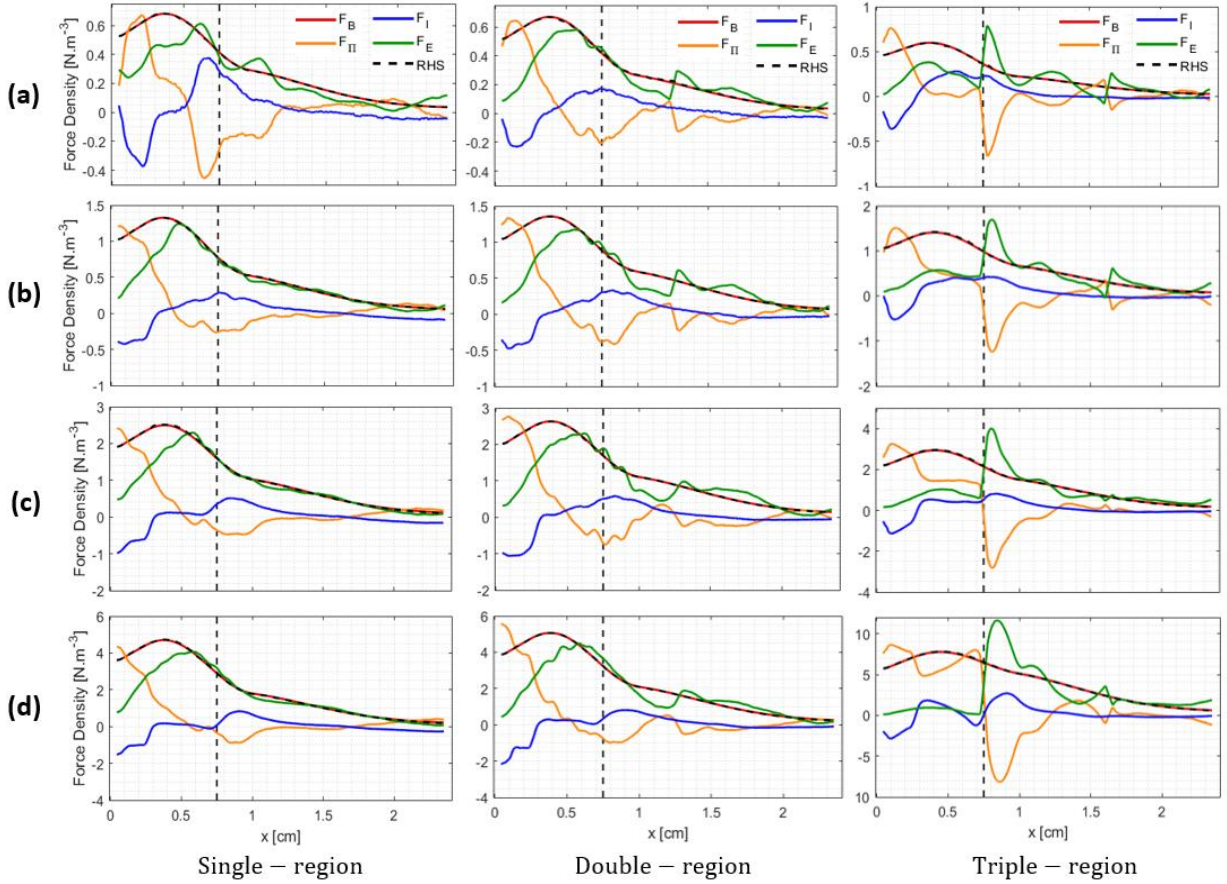

Figure 18: Axial distributions of the force terms in Eq. 23, obtained from the single-region, double-region, and triple-region quasi-2D simulations for different current densities. Rows (a), (b), (c) and (d) correspond to the current densities of 50, 100, 200 and 400 $A/m^2$, respectively. The momentum terms are averaged over the last 4 $\mu s$ of the simulation time.

In Figure 18, we first observe that, in all cases, the sum of all right-hand-side terms, denoted by RHS in the plots and depicted as a dashed black curved line, is perfectly in agreement with the magnetic force term, which implies the satisfaction of the Vlasov equation whose velocity moment yields the momentum equation [28].

Second, as the current density increases from 50 $A/m^2$ (Figure 18, row(a)) to 400 $A/m^2$ (Figure 18, row (d)), the overall magnitude of the magnetic force term increases in all quasi-2D simulations, whereas the location of the $F_B$ peak remains practically invariant. This observation is consistent with that from the 2D reference simulations [18][30].

Third, the magnitude and the distribution of the $F_B$ term for any specific current density case (each row in Figure 18) is almost similar from different quasi-2D simulations. However, the axial distribution of the individual force terms on the RHS of the momentum equation varies for different low-number-of-region quasi-2D simulations, with a noticeable difference observable in the distributions between the double-region and the triple-region simulations. Particularly, concerning the $F_\Pi$ term and the $F_E$ term, which as mentioned above represents the effect of the azimuthal instabilities on electrons' transport, it is noticed that the profiles in the triple-region simulation feature distinct peaks near the location of the maximum magnetic field intensity (shown by vertical dashed lines in Figure 18). The amplitude of the peaks in the profiles of the $F_\Pi$ and the $F_E$ terms decreases when lowering the current density from 400 $A/m^2$ to 50 $A/m^2$.

**Section 5: Verification of the quasi-2D simulations in the limit of high number of regions**

In this section, we present and discuss the results from the quasi-2D axial-azimuthal simulations carried out in the limit of "high" number of regions. The aim is to demonstrate that, in this limit, the quasi-2D results practically converge to the full-2D ones, thus, enabling us to recover the same behaviors and dynamics of the plasma along the axial and azimuthal directions as in a 2D simulation but with several factors reduction in the computational cost.

The setup of the simulations is the same as that described in Section 4.1. The adopted simulation condition, defined in terms of the magnetic field peak intensity and the maximum current density, is the one corresponding to the "benchmark" conditions (Section 4.1), namely, $B = 100\ G$ and $J = 400\ A/m^2$. The "high-order" quasi-2D simulations are performed using various multi-region approximations of the 2D domain, comprising the following cases: case $i$: $N = 5, M = 10$, case $ii$: $N = 10, M = 20$, case $iii$: $N = 20, M = 40$ and case $iv$: $N = 40, M = 80$. In Section A.2 of the appendix, we have additionally reported the results from the quasi-2D simulations with multi-region domain decomposition only along the x-direction, namely, $N = 1, M = 20$ and $N = 1, M = 40$, which are compared against the results from cases $ii$ and $iii$. The computational and physical parameters used for all simulations are the same as those reported in Table 3. The main difference is concerning the initial total number of macroparticles and, hence, the initial macroparticles per cell count. In this regard, to ensure that a minimum of about 100 macroparticles per cell along the azimuthal direction is maintained throughout the simulation in all cases, we chose the initial total number of macroparticles to be 250,000 for case $i$, 500,000 for case $ii$, 1,000,000 for case $iii$, and 2,000,000 for case $iv$. The results from the quasi-2D simulations, in this section, are compared against the results from the 2D simulations with the quasi-neutrality cathode boundary condition, as reported in Ref. [18], in terms of the evolution of the normalized electron and ion currents, the axial profile of the plasma properties, the 2D reconstruction of the plasma properties' distribution, and the dispersion characteristics of the azimuthal instabilities.

**5.1. Time evolution of the electron and ion current**

Figure 19 shows the temporal variation of the normalized electron and ion currents from the quasi-2D simulations. The ion current ($I_i$, Figure 19(b)) corresponds to the flux of the ions leaving the cathode boundary of the domain. Moreover, the difference between the electron and ion flux reaching the anode boundary gives the "discharge" current, whose difference with respect to $I_i$ yields the electron current ($I_e$, Figure 19(a)). In the plots of Figure 19, the y-axis is normalized with respect to the total ion current, i.e., $J \times A$, where $J$ is $400\ A/m^2$ and $A$ is the domain's cross-section area. In addition, the dashed horizontal lines indicate the mean value of the currents, calculated from $t = 5\ \mu s$.

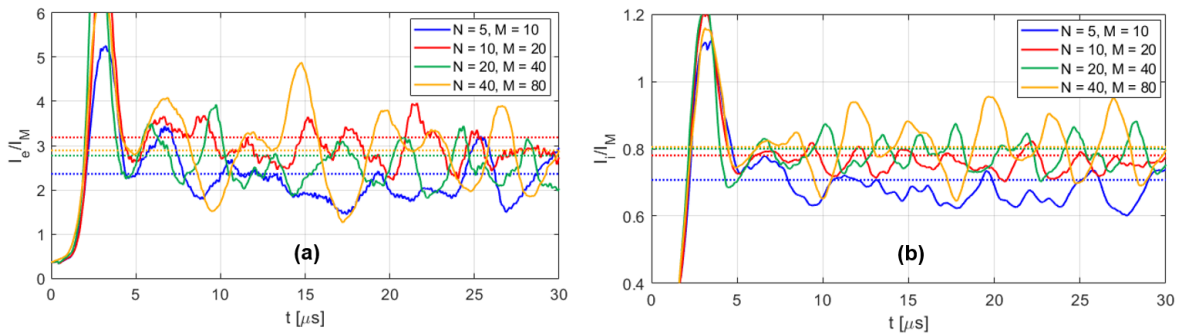

Figure 19: Time evolution of (a) electron current, (b) ion current, from the quasi-2D simulations. The currents are normalized with respect to the nominal total ion current ($I_M$).

Looking at plots in Figure 19, it is observed that, following an initial transient, all simulations have arrived at the quasi-steady state from about $5\ \mu s$. Furthermore, increasing the number of regions from $N = 5, M = 10$ (case $i$) to $N = 40, M = 80$ (case $iv$), the mean values of the normalized electron and ion currents converge to about 3 and 0.8, respectively. These mean current values are very similar to those from the reference 2D simulation [18]. Finally, it is interesting to note that, despite a rather different currents' trace between case $iii$ ($N = 20, M = 40$) and case $iv$, their mean currents are almost the same, implying that these two simulations are expected to provide quite similar predictions in terms of the time-averaged macroscopic plasma properties. We will see in the next section that this is indeed the case.

## 5.2. Distribution of the plasma properties

The axial distributions of the plasma properties from the quasi-2D simulations with different numbers of regions $N$ and $M$ are shown in Figure 20. The 2D reference results in this figure represent those from the full-2D simulation with quasi-neutrality cathode boundary condition [18]. The simulations have been run for 30 $\mu s$, and the profiles are averaged over the last 10 $\mu s$ of the simulation time to be consistent with the 2D reference case.

In the following plots, the axial and azimuthal components of temperature are calculated using the relations

$$T_x = \frac{m}{en}\left(\int_{-\infty}^{\infty} v_x^2 f(\boldsymbol{v})dv_x - n\, V_{d,x}\right), \tag{Eq. 24}$$

$$T_y = \frac{m}{en}\left(\int_{-\infty}^{\infty} v_y^2 f(\boldsymbol{v})dv_y - n\, V_{d,y}\right). \tag{Eq. 25}$$

In the above expressions $v_x$ and $v_y$ are particles' axial and azimuthal velocity, $V_{d,x}$ and $V_{d,y}$ are the axial and azimuthal drift velocities, $f(\boldsymbol{v})$ is the distribution function, $n$ is number density, $e$ is the elementary charge and $m$ is the particle's mass.

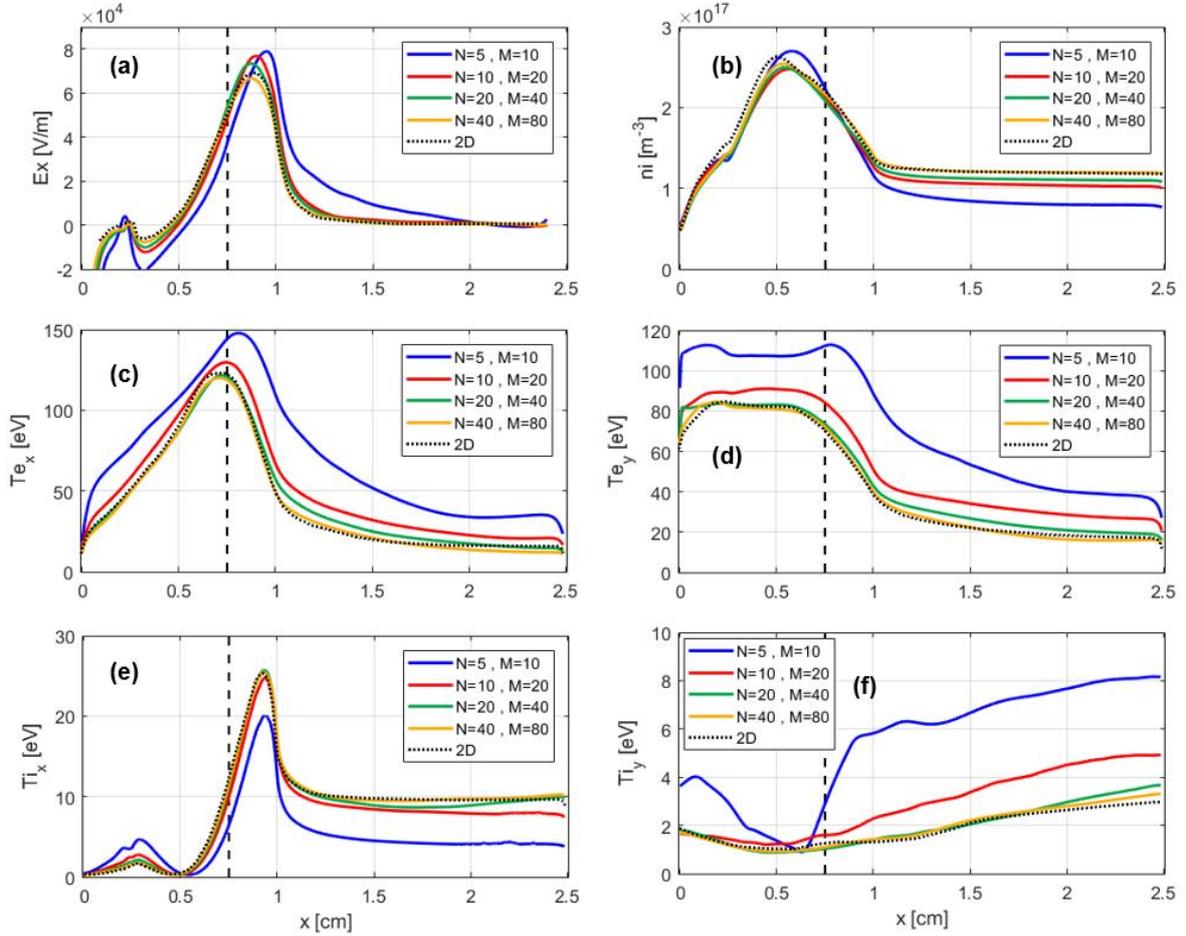

Figure 20: Time-averaged axial profiles of the plasma properties from the "high-order" quasi-2D simulations with different numbers of regions $N$ and $M$; (a) axial electric field, (b) ion number density, (c) and (d) axial and azimuthal electron temperature, (e) and (f) axial and azimuthal ion temperature. The profiles are average over the last 10 $\mu s$ of simulation time. The 2D reference results are from Ref. [18].

The plots in Figure 20 clearly indicate that increasing the number of regions lead to the convergence of the plasma profiles to those from the 2D simulation. In fact, the predictions of case $ii$ ($N = 10$, $M = 20$) is seen to be very similar to those of case $iii$ ($N = 20$, $M = 40$), which remarkably agree with the reference profiles. Moreover, the axial profiles of case $iv$ ($N = 40$, $M = 80$) are almost identical to those for case $iii$ and are in quite a perfect agreement with the 2D results. These observations suggest that, in terms of the axial distribution of plasma properties, increasing $M$ and $N$ further beyond their values for case $ii$ ($N = 10$, $M = 20$) does not significantly change the agreement between the "high-order" quasi-2D and the full-2D simulations' prediction. In fact, the minor difference between the plasma profiles of case $iii$ and case $iv$ with those from the 2D simulation is close to the margin of error observed between various 2D simulations performed by different participating groups in the Hall thruster axial-azimuthal benchmarking activity [15].

Before proceeding further, it is necessary to clarify a point concerning the cathode boundary condition model. In this regard, in Section 4, we showed that "low-order" quasi-2D simulations with quasi-neutrality cathode boundary condition are able to rather accurately reproduce the reference 2D profiles with a current-equality boundary condition. Of course, we will take a close look at the sensitivity of the quasi-2D simulations to the cathode boundary condition in Section 6.1. Nonetheless, for the "high-order" quasi-2D simulation, we demonstrated above the convergence of the quasi-2D results to those from the full-2D simulation with the quasi-neutrality boundary condition [18].

This observation can be justified by noting the way the cathode boundary condition is applied in the quasi-2D simulations. In this regard, we first recall that, in a 2D simulation, the cathode boundary condition is applied to the cells on the cathode plane along the azimuthal direction. Now, in the reduced-order scheme, due to having a quasi-2D grid, we need to apply the cathode boundary condition to the cell(s) on the axial computation grid that are near the location of the cathode plane, which, at the low number-of-region limit, does not coincide with the cells along the azimuthal direction. As we increase the number of regions along the axial direction, the azimuthal grid in the rightmost region of the domain gets closer to the cathode plane and, hence, the non-neutrality on the axial grid at the cathode location captures more closely the average non-neutrality effect along the azimuth at this location in a 2D simulation. Consequently, for the simulations with low number of regions, the quasi-neutrality cathode condition acts more like the current-equality condition in the 2D simulation. However, using larger number of regions, the quasi-2D simulation more closely resembles the full-2D and, accordingly, the applied cathode boundary condition tends to have the same effect as that in the 2D simulation.

Following the above clarification, we now refer to Figure 21, which provides the snapshots of reconstructed 2D distributions of the plasma properties in the $x - y$ plane from the "high-order" quasi-2D simulations at time $t = 30\ \mu s$.

Looking at the reconstructed 2D maps of the ion number density and the azimuthal electric field, it is observed that, although in case $i$ ($N = 5$, $M = 10$), the simulation predicts fluctuations that have only an azimuthal wavevector component, in cases $iii$ and $iv$ (i.e., $N = 20$, $M = 40$ and $N = 40$, $M = 80$, respectively), the $n_i$ and $E_y$ maps clearly show inclined fluctuations before the exit plane, indicated by the dashed vertical lines in Figure 21. This is the evidence that, in the high number-of-region simulation limit, the reduced-order PIC is able to resolve the axial wavevector component ($k_x$) of the azimuthal instabilities. This result points to the remarkable capability of the reduced-order scheme in the high-order approximation limit to capture the underlying physics in the very same manner as a full-2D simulation but at a computational cost reduced by a factor of about 6 in case $iii$ and 3 in case $iv$, whose results are in any case almost identical.

To explain why in cases $iii$ and $iv$ the quasi-2D simulations can resolve the axial wavenumber of the azimuthal instabilities, it is underlined that, should the regions' axial extent be smaller than the waves' axial wavelength, i.e., $l_{x,m} \ll \lambda_x$, the simulations are indeed expected to be able to capture the axial component of the wavevector. This criterion is met in both cases $iii$ and $iv$. For instance, focusing on case $iii$, as it can be noticed from the $x - y$ fluctuation map of the azimuthal electric field in Figure 21, the azimuthal wavelength ($\lambda_y$) of the waves is about 1 mm. Now, we recall that, according to the literature [29], the instabilities' wavelength along the axial direction is much larger than that along the azimuthal direction ($\lambda_x \gg \lambda_y$). Therefore, considering that the axial length of the vertical regions ($l_{x,m}$) in case $iii$ is 0.625 mm, we have $l_{x,m} < \lambda_y$ and $\lambda_y \ll \lambda_x$, thus, $l_{x,m} < \lambda_x$. This same analysis is valid for case $iv$, where the length of the vertical regions is smaller than that in case $iii$ but the waves' azimuthal wavelength is almost the same.

Accordingly, the reconstructed 2D fluctuation maps for cases $iii$ and $iv$ (the third and fourth row in Figure 21) were found to be in remarkable agreement with the similar maps from the 2D simulation in Ref. [18] in terms of the azimuthal wavelength and the tilt in the waves before the channel exit. Moreover, in the 2D simulation, the instabilities were observed to become almost purely azimuthal downstream of the channel exit plane, which is also seen in the fluctuation maps of cases $iii$ and $iv$.

Finally, concerning case *ii* ($N = 10$, $M = 20$), we notice that the axial wavenumber of the azimuthal waves is partially captured but the axial extent of the vertical regions has not been small enough in this simulation case to allow for an accurate resolution of $k_x$.

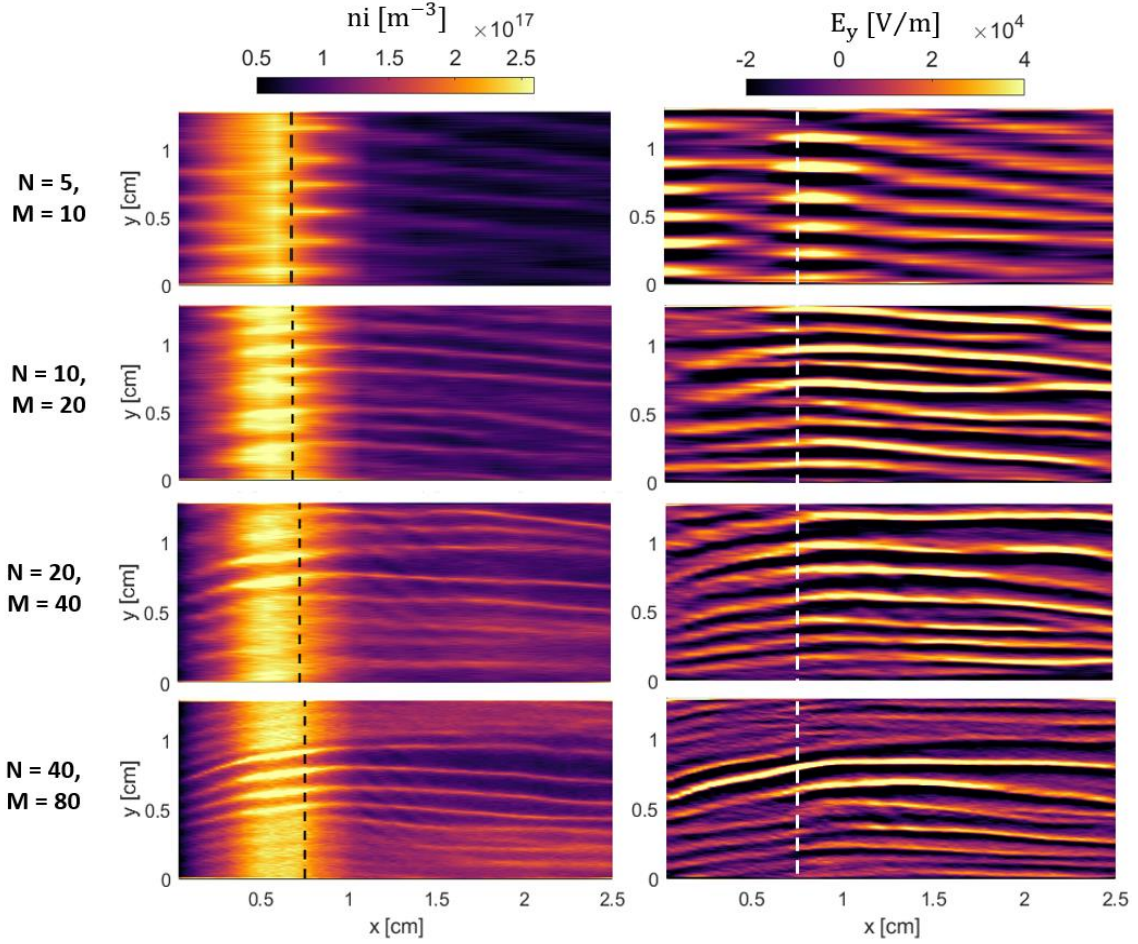

Figure 21: Reconstructed 2D maps of the ion number density (left column) and the azimuthal electric field (right column) from "high-order" quasi-2D simulations with different numbers of regions $N$ and $M$. The dashed lines show the location of the peak magnetic field intensity.

Figure 22 illustrates the dispersion plots of the azimuthal electric field fluctuations at three axial locations from the "high-order" quasi-2D simulations. These axial locations are chosen such that they correspond as closely as possible to the mid-plane locations of the three regions in the triple-region simulation discussed in Section 4.3.

Looking at the dispersion plots, all "high-order" simulation cases show a minor variation in the loci of dominant frequencies and wavenumbers from the axial location of 0.4 cm to 2 cm. The variation in the wavenumbers is mostly from larger to smaller values when moving from inside the channel ($x = 0.4\ cm$) toward the near-plume. In addition, the waves show a broader dispersion in the near-plume region. Also, as pointed out in Section 4.3 for the low-order quasi-2D simulations, the increasing deviation seen here between the local ion acoustic dispersion relation and the one corresponding to the upstream region (whose mid-plane falls at $x = 0.4\ cm$) is again the evidence of the downstream convection of the azimuthal waves by the ion beam.

Finally, comparing the dispersion plots in Figure 22 with those in Figure 17(d) for the triple-region simulation in the benchmark condition, it can be noticed that the prominent characteristics of the azimuthal waves observed from the "high-order" quasi-2D simulations have been already captured to a great extent in the triple-region simulation. This indicates that even lower order approximations of the 2D problem could allow us to resolve the underlying azimuthal physics with sufficient accuracy.

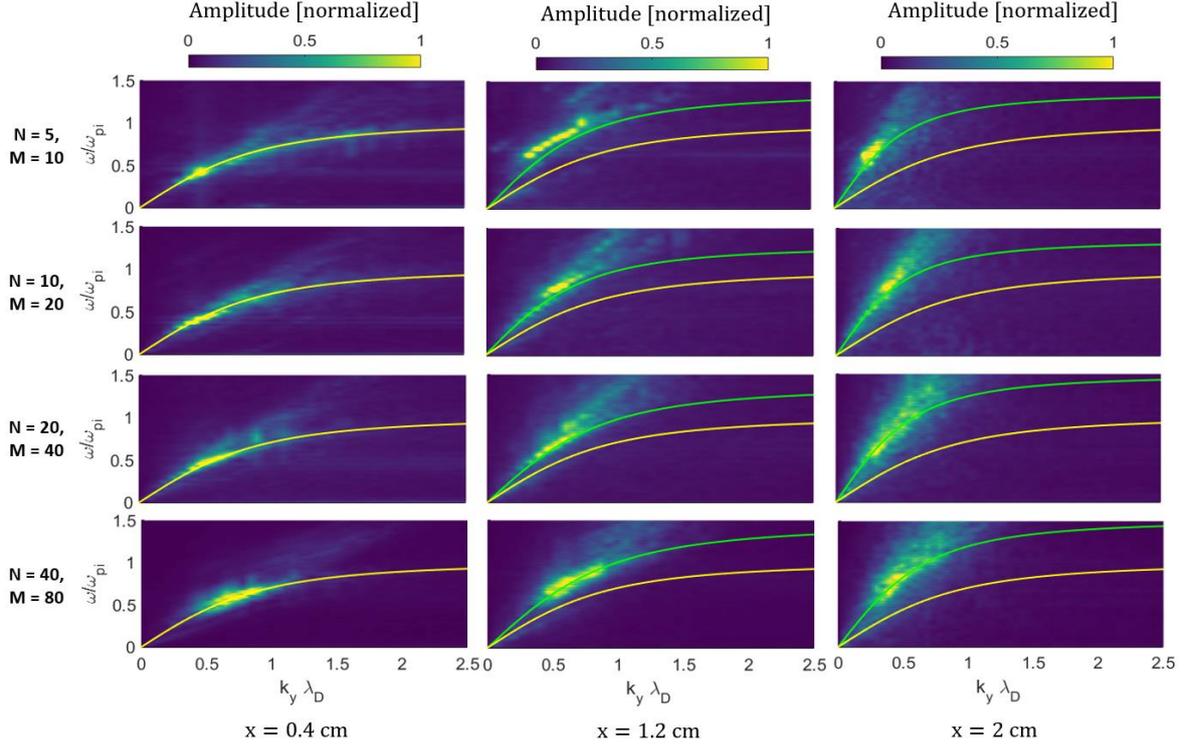

Figure 22: Dispersion plots of the azimuthal electric field oscillations at the axial location of 0.4 cm (left column), 1.2 cm (middle column) and 2 cm (right column) from the quasi-2D simulations with different numbers of regions $N$ and $M$. Yellow lines are the local ion acoustic dispersion relations, and the green lines are the dispersion relations corresponding to the vertical region whose mid-plane falls at $x = 0.4$.

### Section 6: Analysis of the sensitivity of the reduced-order axial-azimuthal simulations

In this section, we present the results from a series of quasi-2D simulations that we carried out to evaluate the sensitivity of the reduced-order scheme to the main numerical aspects of the axial-azimuthal simulations. In all cases, the simulation conditions correspond to those of the "benchmark" [15] (Section 4.1).

**6.1. Effect of the cathode boundary condition**

As mentioned previously, the quasi-2D simulations discussed so far are performed using the quasi-neutrality boundary condition. We have found that the reduced-order scheme behaves more stably with this cathode boundary condition in the axial-azimuthal simulation case adopted for this work. Nevertheless, to assess more rigorously the impact of the cathode boundary condition on the quasi-2D simulations' results, we have performed single-region and double-region simulations with both the quasi-neutrality (QN) and the current-equality (CE) cathode models and have compared the results in terms of the axial profiles of the time-averaged plasma properties as well as the evolution and characteristics of the azimuthal instabilities.

Before presenting the results, however, we present in the following the main implementation details for the two cathode boundary conditions mentioned above.

A – **Current equality between the current reaching the anode and the cathode plane (CE)**: In this model, we keep track of the number of electrons and ions reaching the anode and inject electrons from the cathode boundary at a rate that satisfies the current continuity throughout the electric circuit. This implies that the sum of the ion and electron currents collected on the anode surface (right boundary of the computational domain) must be equal to that reaching the cathode plane (left boundary of the computational domain). The electrons' injection model based on the current-equality condition is implemented in the quasi-2D simulations according to the description provided in Ref. [18], where the cathode plane is considered to be 1 mm inwards from the right-end boundary of the domain. The flux of the electrons to be injected from this cathode location ($\Gamma_{ce}$) at each timestep is obtained as

$$\Gamma_{ce} = \Gamma_{ae} - \Gamma_{ai}, \tag{Eq. 26}$$

where $\Gamma_{ai}$ and $\Gamma_{ae}$ are ion and electron fluxes reaching the anode.

B – **Quasi-neutrality on the cathode plane (QN)**: Alternatively, we can inject from the cathode as many electrons as necessary to keep the cathode plane quasi-neutral at each timestep. The number of electrons to be injected ($N_e$) corresponds to the difference between ion and electron number densities on the cathode plane which, in 2D, is obtained as

$$N_e = \sum_{j=1}^{N_j} \left(n_{i_j} - n_{e_j}\right)|_{x=x_c} \Delta V, \quad \text{(Eq. 27)}$$

where $n_i$ and $n_e$ are ion and electron number densities, $N_j$ is the number of cells along the azimuthal directions, $\Delta V$ is the cell volume, and $x_c$ is the cathode axial location. According to the QN cathode boundary model, we only inject electrons if $N_e > 0$.

In the quasi-2D case with a single region along the azimuthal direction ($N = 1$), the above condition is applied to the cell on the axial computation grid that is at the cathode axial location; hence, we have

$$N_e = (n_i - n_e)|_{x=x_c} \Delta V, \quad \text{(Eq. 28)}$$

The above relation is extended to the quasi-2D simulations with multiple regions $N$ along the azimuthal direction using the following equation

$$N_e = \sum_{nr=1}^{N} (n_i^{nr} - n_e^{nr})|_{x=x_c} \Delta V^{nr}. \quad \text{(Eq. 29)}$$

In Eq. 29, the superscript $nr$ denotes each specific horizontal region along the azimuth, and the terms with $nr$ superscript are the same quantities defined in Eq. 27 but corresponding to their values in each horizontal region. The QN method is shown to be less sensitive to the electrons' injection temperature [18], which is unknown without modelling self-consistently the cathode plasma and its coupling with the thruster main plume.

The axial-azimuthal simulations are typically sensitive to the way in which the cathode boundary condition is implemented [18]. In this regard, the QN model is more suitable for the transient phase before reaching the quasi-steady state. This is because the charge imbalance introduced as a result of the CE boundary condition during the simulation's transient regime may disturb the charge distribution and, consequently, the waves' structure at the cathode plane, which can in turn affect the simulation results at the steady state [18].

Following the above descriptions of the cathode models, we now refer to Figure 23, which shows that all plasma profiles from our "low-order" quasi-2D simulations are influenced by the cathode boundary condition. In particular, the simulations with the CE boundary condition, especially the single-region case, predict a higher plasma density peak compared to the simulations with the QN boundary. In addition, the peak of the electric field is higher when using the CE condition and the region of maximum electric field extends over a narrower extent. Finally, the CE boundary condition is observed to result in a rise in the electron temperature near the cathode boundary. However, in the case of the QN cathode injection model, the electron temperature is maintained at the value at which the electrons are injected into the domain from the cathode plane (i.e., 10 eV).

Concerning the quasi-2D simulations with high number of regions, we have noticed that the sensitivity of the simulations to the cathode injection model becomes more pronounced, and the CE cathode boundary condition leads to further deviation of the results from with respect to the reference simulations. Whereas as it is shown in Section 5, the predictions of the "high-order" quasi-2D simulations with the QN condition indeed converge to the 2D results.

The comparison between the characteristics of the azimuthal waves in the simulations with different cathode boundary conditions is presented in Figure 24 and Figure 25. From these figures, it is noticed that the waves developed in the single-region simulations with the QN and CE boundary conditions share similar properties in terms of both the amplitude and the dispersion characteristics. However, in the double-region simulations with different cathode boundary conditions, although the waves in Region I feature similar spatiotemporal behavior and dispersion characteristics between the two cases, the azimuthal waves in Region II, in which the effect of the cathode boundary condition is mostly reflected, are very different. In case of the CE condition, the instabilities' dispersion is discrete, representing several distinct dominant modes, whereas the QN condition has led to a more continuum dispersion of the waves.

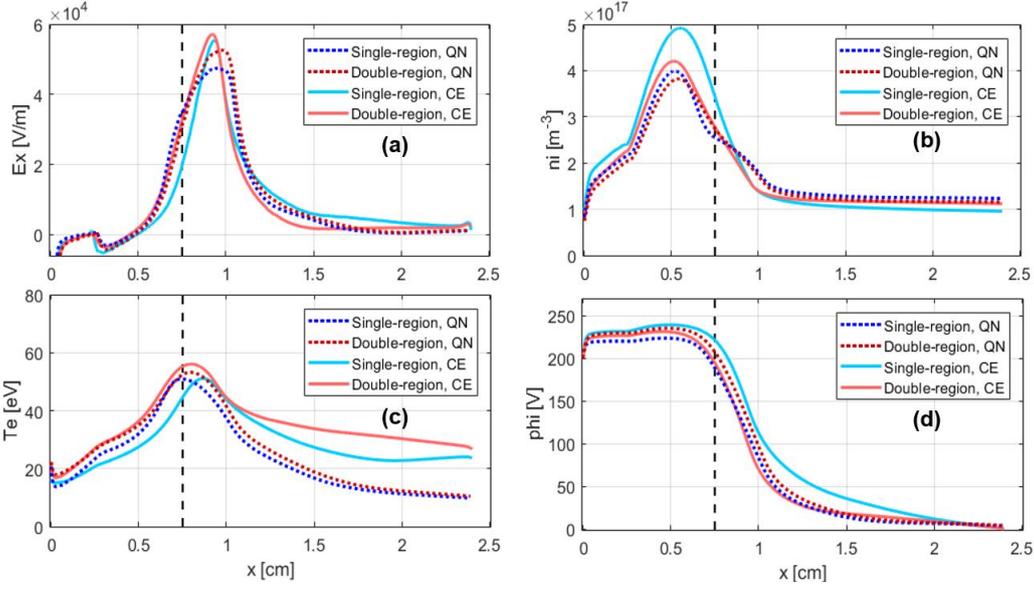

Figure 23: Comparison between the time-averaged axial profiles of plasma properties from the single- and double-region simulations with quasi-neutrality (QN) and current-equality (CE) cathode boundary conditions; (a) axial electric field, (b) ion number density, (c) electron temperature and (d) electric potential.

The above observations regarding the waves' characteristics in Region I and II of the double-region simulations with the QN and CE boundary conditions clarify the reason why, in the plots of Figure 23, the ion number density and electron temperature distributions from the CE simulation are relatively similar in the upstream (Region I) to those from the QN simulation, but that the electron temperature is, in particular, noticeably different between the two cases of the double-region simulation in the near-plume (Region II).

Before concluding the discussion on the sensitivity of the quasi-2D simulations to the cathode boundary condition, it is underlined that, because of the presence of a predefined ionization source in the adopted reference simulation case for the present study, the difference between the results obtained with different cathode boundary conditions might be exaggerated with respect to when the ionization process is captured self-consistently. In fact, this point is also acknowledged in the 2D reference publication [18].

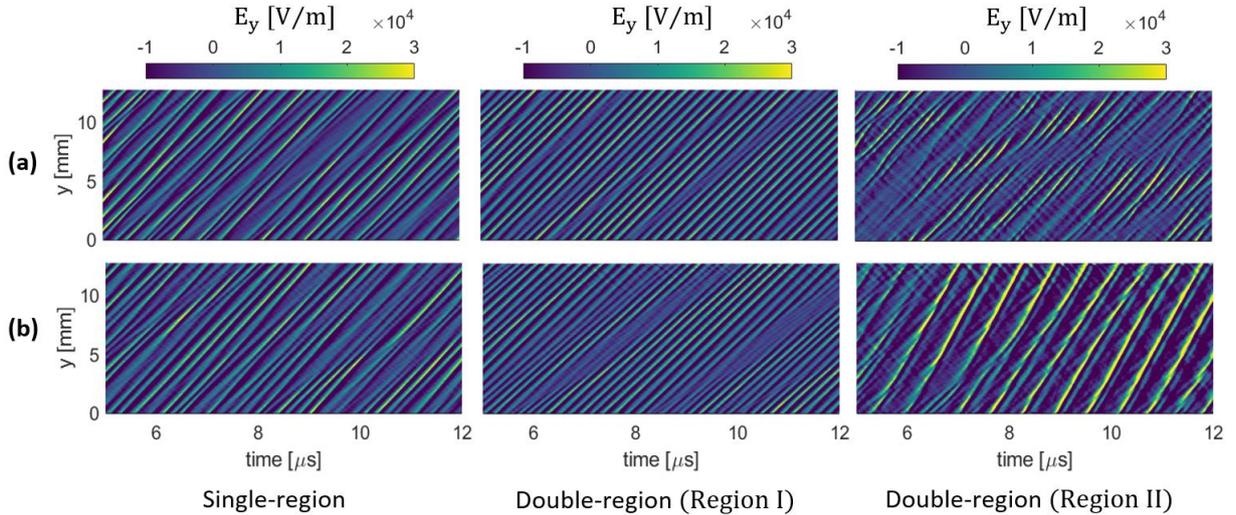

Figure 24: Spatiotemporal maps of the azimuthal electric field fluctuations from the single-region simulation (left column), and the Region I (middle column) and Region II (right column) of the double-region simulation. Rows (a) and (b) correspond, respectively, to simulations with the quasi-neutrality (QN) and current-equality (CE) cathode boundary conditions.

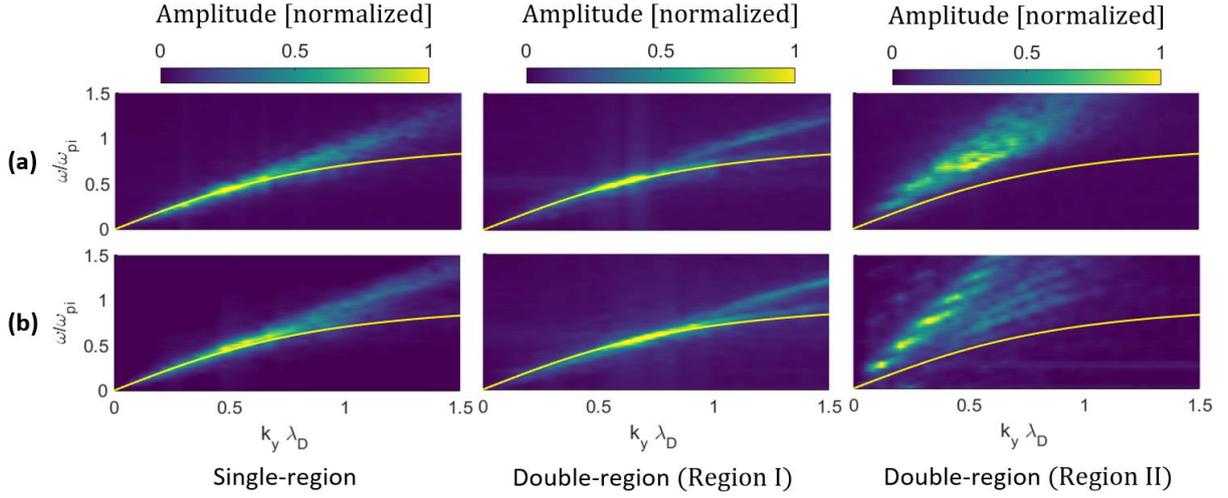

Figure 25: Dispersion plots of the azimuthal electric field fluctuations from the single-region simulation (left column), and the Region I (middle column) and Region II (right column) of the double-region simulation. Rows (a) and (b) correspond, respectively, to simulations with the quasi-neutrality (QN) and current-equality (CE) cathode boundary conditions.

### 6.2. Effect of the azimuthal extent of the domain

As simulating the entire circumference of a Hall thruster has been computationally impractical in conventional full-2D simulations, it has become customary to instead simulate a limited portion of the circumferential extent and apply a periodic boundary condition on both ends of the domain along the azimuthal direction. Although the impact of the periodicity assumption on the waves' characteristics and on the simulation as a whole without any scaling of the physical parameters and for real-world simulation conditions is still unknown, the recent 2D simulations of the reference case adopted in this work suggest minor sensitivity of the results to the simulated azimuthal length [18][30]. Accordingly, the effect of the azimuthal extent of the domain is another aspect with respect to which the reduced-order PIC scheme must be verified. In this regard, we showed in our previous work [10] that the single-region simulation with the preliminary dimensionality-reduction formulation did not have a significant sensitivity to the azimuthal domain length. Here, we present a comparison between the results from the triple-region simulations with the mature dimensionality-reduction formulation (see Section 2.1) for azimuthal lengths of 0.64, 1.28 and 2.56 cm.

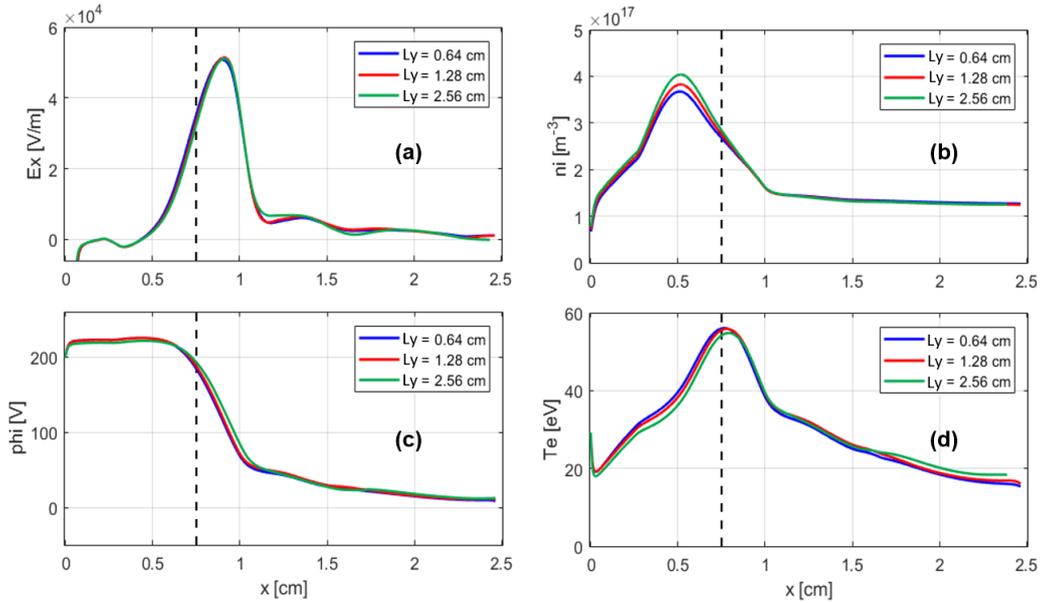

Figure 26: Comparison between the axial profiles of the time-averaged plasma properties from the triple-region simulations with various azimuthal extents; (a) axial electric field, (b) ion number density, (c) electric potential and (d) electron temperature

Concerning the axial distribution of the time-averaged plasma parameters, Figure 26 illustrates a very minor sensitivity of the predicted profiles to the azimuthal length of the domain. The observed sensitivity of the quasi-2D simulation is found to be very similar to the sensitivity to the azimuthal length reported for the full-2D simulations [18].

Figure 27 shows the spatiotemporal maps of the azimuthal electric field fluctuations in each region of the triple-region simulations for various domain's azimuthal extents. Similar to what was observed from the axial profiles of the plasma properties, we notice that the waves' temporal behavior and amplitude in each region are almost the same for all simulated azimuthal domain lengths. This observation could have been expected since the great extent of similarity between the time-averaged plasma properties' profiles in Figure 26 was indicating that the effect of the azimuthal instabilities on electron transport, which has a dominant role in determining the plasma distributions and is related mostly to the amplitude of the oscillations, has been almost the same across all cases.

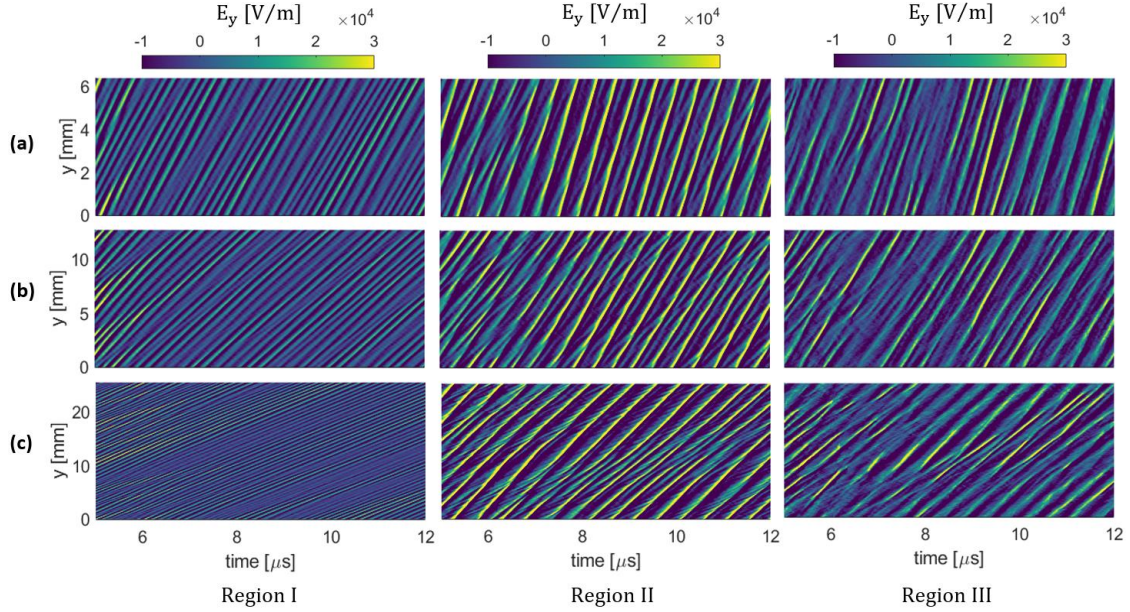

Figure 27: Spatiotemporal maps of the azimuthal electric field fluctuations in Region I (left column), Region II (middle column), and Region III (third column) of the triple-region simulations with the azimuthal lengths of (a) 0.64 cm, (b) 1.28 cm, and (c) 2.56 cm.

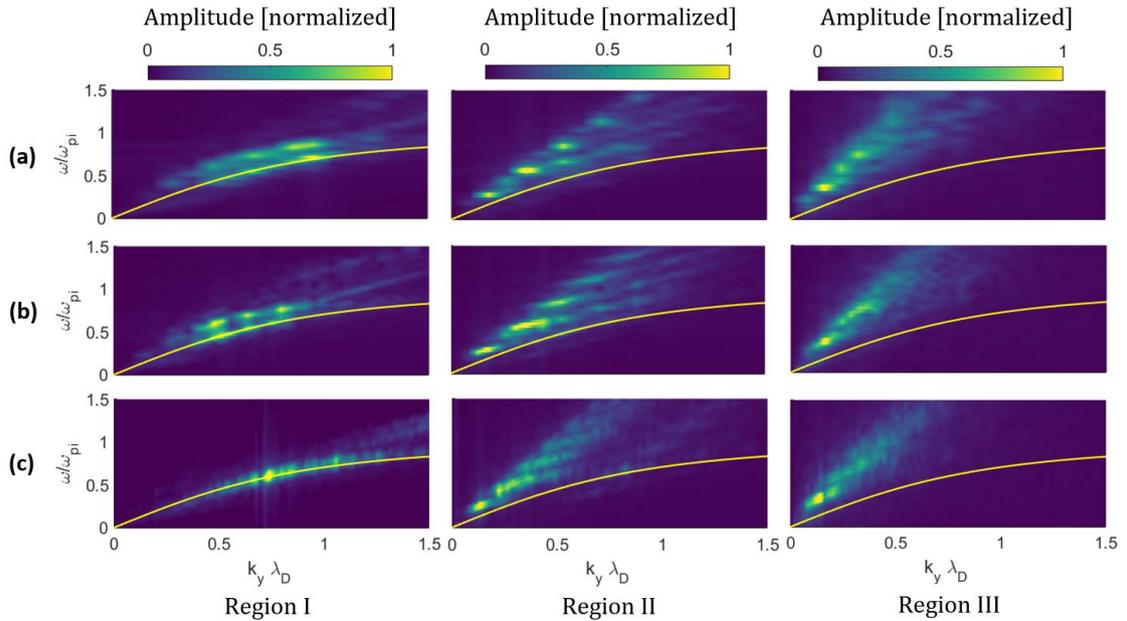

Figure 28: Dispersion plots of the azimuthal electric field fluctuations in Region I (left column), Region II (middle column), and Region III (right column) of the triple-region simulations with the azimuthal lengths of (a) 0.64 cm, (b) 1.28 cm, and (c) 2.56 cm.

The dispersion characteristics of the azimuthal instabilities are also seen from Figure 28 to be relatively similar among the simulation cases with various lengths of the azimuthal extent. In any case, some variation in the dispersion features of the waves can be noticed. In this regard, the waves in all regions of the triple-region simulation with the azimuthal length of 2.56 cm (Figure 28, row (c)) show a dominant wave mode. Also, we can see less dispersion in the waves' frequency in Region I of the simulation with $L_y = 2.56$ cm. In the other two simulation cases, we first see, in Region I, a broader dispersion in the waves' frequency and almost no particularly dominant wave mode. Second, in Region II, several dominant wave modes spreading over a range of frequencies and wavenumbers are observed. Nevertheless, concerning Region III, the waves' dispersion characteristics from different simulation cases show the most similarity.

### 6.3. Effect of the number of macroparticles per cell in the "low" and "high" number-of-region limits

Another important numerical aspect, whose impact on the quasi-2D simulation results needs to be evaluated, is the number of macroparticles per cell, which is a measure of the statistical quality of sampling of the local distribution function in each computational cell. In this section, we present the results from both "low-order" and "high-order" quasi-2D simulations with various numbers of macroparticles per cell to assess separately the effect of this parameter on the reduced-order simulation predictions at each limit. For the assessment of the sensitivity of the "low-order" simulations, we have chosen the single-region case whose predictions were in quite good agreement with the 2D reference results and also has the lowest computational cost. Concerning the sensitivity in the high number-of-region limit, we chose the multi-region simulation case with $N = 20$ and $M = 40$ because we saw its predictions to be almost the same as the full-2D results and is computationally less expensive than the cases with higher number of regions.

Starting with the single-region simulation, Figure 29 presents the axial profiles of the time-averaged plasma properties from the simulations with varying initial numbers of macroparticles per cell along the azimuthal direction ($N_{ppc_y}$). We have changed this parameter from the nominal value of 300 to both lower values, 75 and 150, and to higher values, 600, 1200, and 2400. It is observed from the plots in Figure 29 that the simulations are indeed sensitive to the value of $N_{ppc_y}$. However, from 300 to 2400 macroparticles per cell, the sensitivity is rather moderate. In addition, the simulation cases with 1200 and 2400 macroparticles per cell have almost identical predictions.

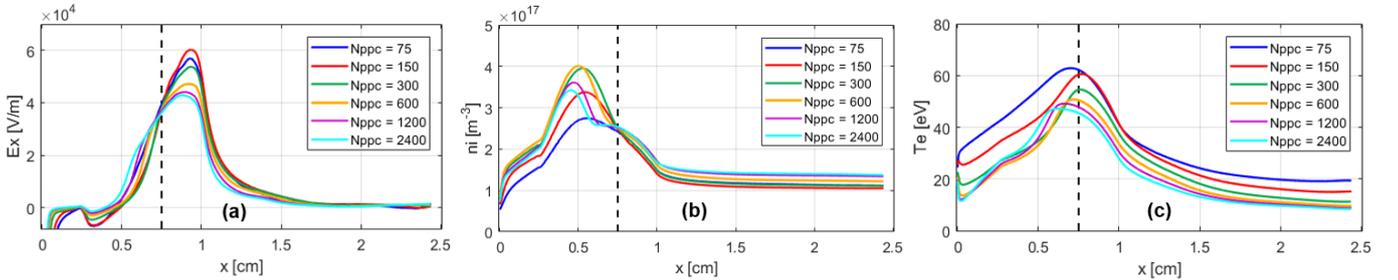

Figure 29: Comparison between the axial profiles of the time-averaged plasma properties from the single-region simulations with various initial numbers of macroparticles per cell; (a) axial electric field, (b) ion number density, and (c) electron temperature

Looking now at the plots in Figure 30, which show the axial profiles of the time-averaged plasma properties from "high-order" quasi-2D simulations with different $N_{ppc_y}$ values, the degree of sensitivity is completely different compared to the "low-order" simulations. In the "high-order" limit, we carried out simulations with the initial number of macroparticles per cell along the azimuthal direction varying from 100 (i.e., the nominal value) to 50, and from 100 to 200 and 400. It is clearly noticed that the results are only marginally sensitive to the value of the parameter $N_{ppc_y}$. Additionally, the observed sensitivity of the "high-order" simulations was found to be the same as the sensitivity of the 2D reference simulation [15] to the $N_{ppc_y}$ value.

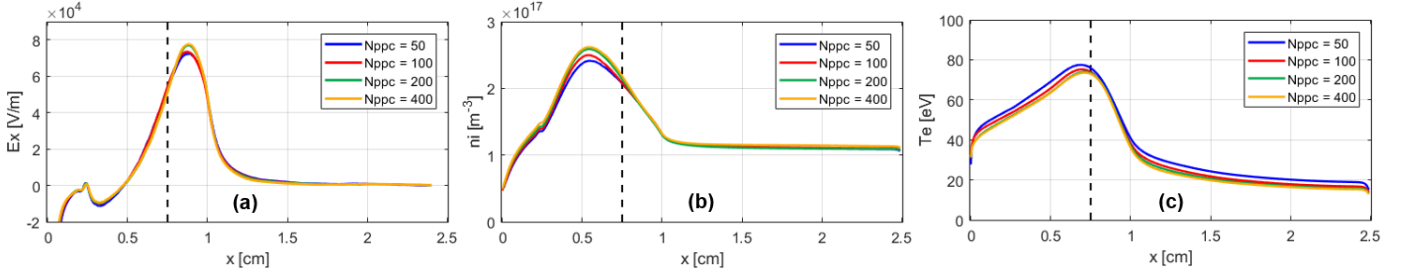

Figure 30: Comparison between the axial profiles of the time-averaged plasma properties from the "high-order" ($N = 20, M = 40$) quasi-2D simulations with various initial numbers of macroparticles per cell; (a) axial electric field, (b) ion number density, and (c) electron temperature.

The difference in the sensitivity of the quasi-2D simulations to the initial number of macroparticles per cell between the low and high number-of-region cases is due to the disparity between the weight of the simulation macroparticles in the two limits. In the single-region simulation, for instance, as there is only one azimuthal grid, having 300 macroparticles per cell corresponds to a total initial macroparticles count of about 75000. However, in the multi-region simulation with $N = 20$ and $M = 40$, as we have 40 azimuthal grids, an initial $N_{ppc_y}$ of 100 corresponds to the initial total macroparticles count of about 1,000,000. Consequently, the macroparticle weight in the "high-order" multi-region simulation is significantly smaller than that in the "low-order" single-region simulation. Accordingly, the charge imbalance, e.g., that injecting particles due to the ionization or from the cathode has on the "low-order" simulations is more prominent than that in the "high-order" simulations. As a result, the sensitivity to the number of macroparticles is also more noticeable in the "low-order" simulations than in the "high-order" ones.

**Section 7: Conclusions**

In this article, we presented a mathematically rigorous dimensionality-reduction technique for Poisson's equation that enables a generalized reduced-order PIC scheme. It was underlined that the main difference between this novel scheme and a conventional PIC is related to the calculation of the electric potential using a reduced-dimension Poisson solver. In this respect, as the validity of the underlying formulation of the dimensionality-reduction was mathematically verified and the generalizability of the reduced-dimension Poisson solver was demonstrated in several standalone Poisson problems, the resulting reduced-order PIC scheme has, in principle, a broad applicability as well. Therefore, this effort significantly advances the concept of the "pseudo-2D" PIC, introduced in our previous publication [10], by addressing the shortcomings that we had identified for preliminary pseudo-2D scheme, namely, its lack of a well-founded formulation for the splitting of Poisson's equation, and the very limited generalizability either with respect to the order of approximation of the 2D problem or applicability to various simulation cases and physical configurations. Consequently, the concept of the reduced-order PIC scheme and its numerical implementation is demonstrated in this work to be mature enough to allow progress toward a computationally efficient, quasi-3D simulation tool for the study of the physics in Hall thrusters and potentially other low-temperature plasma systems.

Indeed, we showed in this paper that, in a 2D axial-azimuthal simulation configuration and compared to the results from a reference 2D case in this configuration, "low-order" quasi-2D simulations, which offer about two orders of magnitude reduction in the computational cost, can overall provide a remarkably close prediction of the time-averaged plasma properties over a broad range of simulation conditions (various current densities and magnetic field peak intensities) with as few numbers of regions as two, i.e., the double-region simulation. Furthermore, with respect to the underlying azimuthal physics and coupling to the axial processes such as cross-field electron mobility, we demonstrated that, on the one hand, the single-region simulation provides a reasonably accurate, average representation of the azimuthal instabilities and captures their overall effect on transport in a manner similar to 2D simulations. On the other hand, "low-order" multi-region simulations resolve several additional aspects of the azimuthal waves' characteristics and evolution, such as the transition in waves' properties around the ion sonic point [15][30] and downstream convection of the waves excited upstream of the domain's exit plane by the accelerating ion beam [30]. At the "low-order" multi-region approximation limit, it was also shown that the simulations are rather insensitive to the length of the domain along the azimuthal direction but exhibit a relatively notable sensitivity to the implementation of the cathode boundary condition.

Furthermore, we demonstrated that, as we increase the order of approximation of the 2D problem by increasing the number of regions, the predictions of the quasi-2D simulations converge to those from the 2D reference simulation, reproducing the time-averaged plasma properties with less than about 2% error and recovering the same features and characteristics observed for the azimuthal waves and their 2D distribution, at an approximation limit that still offers about a factor of 6 reduction in computational cost. Moreover, it was highlighted that even the sensitivity of the "high-order" quasi-2D simulations to the cathode boundary condition model and the number of macroparticles per cell is similar to that observed for the 2D codes.

Accordingly, the results and analyses shown in this paper first confirm that the observations presented in the previous publication using the preliminary pseudo-2D scheme can be reproduced by the generalized reduced-order PIC that features a mature formulation of the dimensionality-reduction technique. Second, the discussions complement the analyses presented in Ref. [10] by extending the investigations to a broader range of simulation conditions using various approximation orders of the 2D problem. As such, the present effort serves as concluding the verifications of reduced-order PIC scheme in the axial-azimuthal configuration. Accordingly, as the future work, we are continuing the verifications in azimuthal-radial and axial-radial configurations of Hall thrusters, on path to the realization of a reliable quasi-3D kinetic simulation code. Also, we intend to define benchmark cases representative of other low-temperature plasma configurations to verify the capabilities of the reduced-order scheme in a wider range of physical setups and applications.

**Acknowledgments:**

The present research is carried out within the framework of the project "Advanced Space Propulsion for Innovative Realization of space Exploration (ASPIRE)". ASPIRE has received funding from the European Union's Horizon 2020 Research and Innovation Programme under the Grant Agreement No. 101004366. The views expressed herein can in no way be taken as to reflect an official opinion of the Commission of the European Union. The authors also acknowledge the computational resources and support provided by the Imperial College Research Computing Service (http://doi.org/10.14469/hpc/2232).

**Appendix**

**A.1. Sensitivity of the reduced-order PIC scheme to the size of the moving-average window in the reduced-dimension Poisson solver**

As described in detail in Section 2.2, the implementation of the reduced-dimension Poisson solver in the PIC code involves modifying slightly the coupled system of 1D ODEs given by Eqs. 6 and 7 such that some of the cross-derivative terms are obtained from the previous timestep of the simulation. It was mentioned in Section 2.2 that, before calculating these derivative terms at each timestep, we applied a moving-average filter with a certain window size on the potential solutions from the previous timestep. Accordingly, it is important to assess the impact of the size of the adopted window on the quasi-2D simulations to ensure that this numerical solver parameter does not have a major influence on the results.

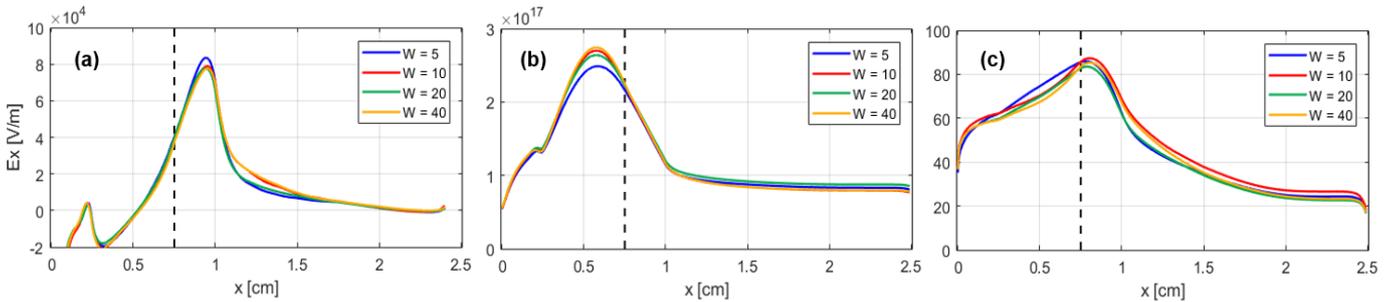

Figure 31: Comparison between the axial profiles of the time-averaged plasma properties from "high-order" ($N = 5, M = 10$) quasi-2D simulations with different moving-average window sizes in the potential solver; (a) axial electric field, (b) ion number density and (c) electron temperature.

In this regard, Figure 31 shows the axial profiles of the time-averaged plasma properties from the multi-region ($N = 5, M = 10$) quasi-2D simulations for various moving-average window sizes. It is observed that the plasma distributions are very closely comparable for all window sizes, with the plasma profiles from the simulations with the window size of 10 (the nominal value), 20 and 40 being almost identical. Of course, the simulation case with the window size of 5 predicts a minimally higher electric field peak and a slightly lower ion number density peak, but its predicted profiles are overall in line with the other cases. As a result, we conclude that the reduced-order simulations are almost insensitive to the moving-average filter's window size, particularly when the size of the window is equal to or great than 10. In a more general sense, the results from the reduced-order simulations are not affected by the window size once the application of the moving-average filter effectively smooths out the noise on the potential solutions.

## A.2. Quasi-2D simulation results with single vs. multiple regions along the azimuthal direction

The current understanding of the plasma behavior in the axial-azimuthal configuration of Hall thrusters indicates that the axial electric field does not have a notable gradient along the azimuthal direction. As a result, one may wonder if it might be actually necessary to divide the azimuthal coordinate into multiple regions in the reduced-order axial-azimuthal simulations, since such a division would be essentially intended to resolve the azimuthal gradients of the axial electric field. The results and analysis here are aimed at answering this question.

To this end, we have carried out dedicated "high-order" multi-region quasi-2D simulations with only a single horizontal region along the azimuth. These new simulation cases have domain decompositions with $N = 1, M = 20$ and $N = 1, M = 40$, and their results are compared against the corresponding simulations with $N = 10, M = 20$ and $N = 20, M = 40$. In this regard, Figure 32 presents the comparison between the axial profiles of the plasma properties from the simulation cases with single and multiple decomposition of the azimuthal coordinate. We observe that the profiles from the simulations with $N = 1$ are marginally different from the corresponding profiles from the simulations with $N = 10$ and 20.

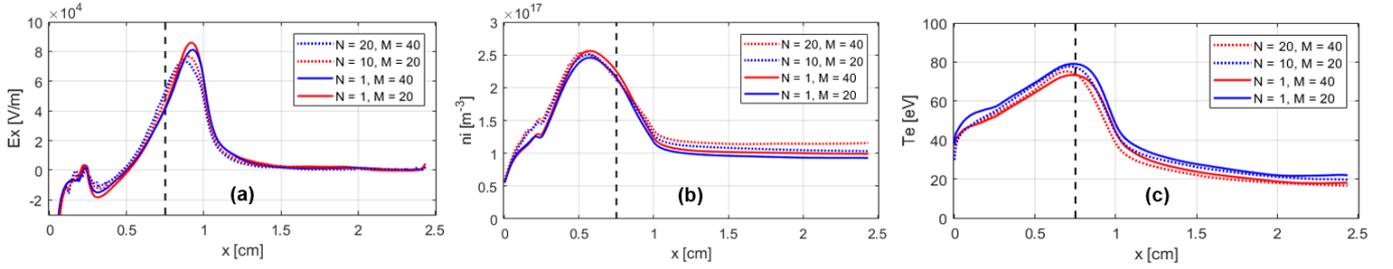

Figure 32: Comparison between the axial profiles of the time-averaged plasma properties from "high-order" quasi-2D simulations with different numbers of horizontal regions ($N$) along the azimuthal coordinate; (a) axial electric field, (b) ion number density and (c) electron temperature.

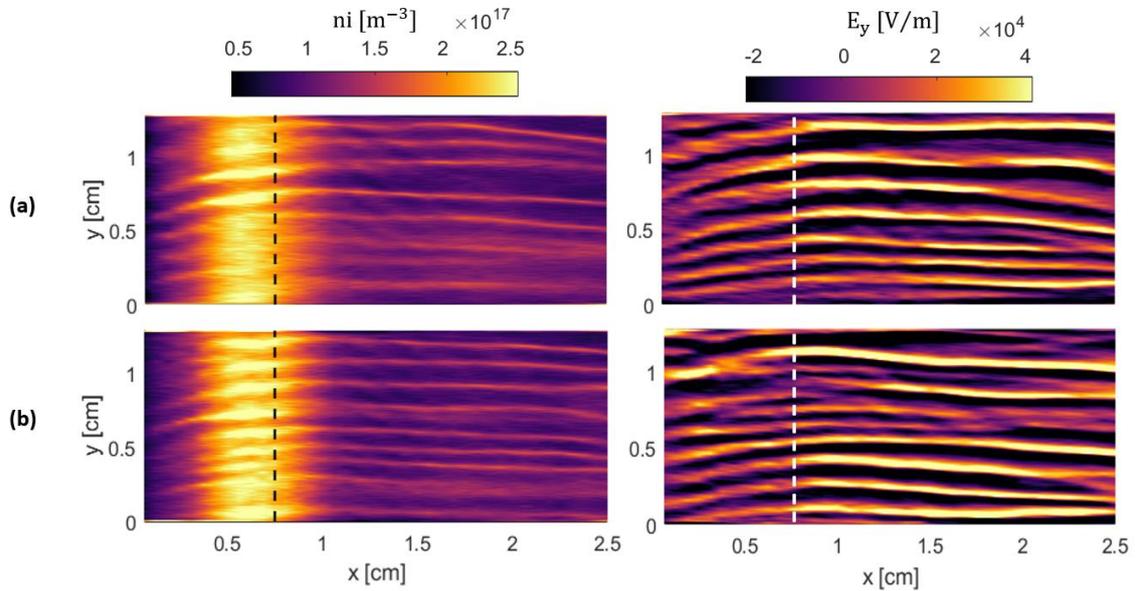

Figure 33: Reconstructed 2D maps of the ion number density (left column) and the azimuthal electric field (right column) from "high-order" quasi-2D simulations with different numbers of horizontal regions ($N$) along the azimuthal coordinate; (a) $N = 20, M = 40$, (b) $N = 1, M = 40$. The dashed lines show the location of the peak magnetic field intensity.

Moreover, in Figure 33, we have compared the reconstructed 2D snapshots of the ion number density and azimuthal electric field at $t = 30\ \mu s$ from the simulations with $N = 20, M = 40$ (Figure 33, row (a)) and $N = 1, M = 40$ (Figure 33, row (b)). We notice that, in terms of the reconstructed 2D maps of the plasma properties as well, the results from the two simulations do not show any major difference, with the $N = 1$ simulation capturing all main features of the reconstructed distributions from the $N = 20$ simulation.

The above results underline the fact that, as initially claimed, the gradient of the axial electric field along the azimuthal direction is rather negligible, and a simulation that does not resolve this gradient provides predictions that compare very well with the simulations that do resolve this aspect. Consequently, we can state that, in general, an a-priori understanding of the physics of a specific problem can inform the decision concerning the degree of domain decomposition (i.e., the number of regions) such that the computational gain from the reduced-order scheme could be maximized.

**References**:

[1] Boeuf JP, "Tutorial: Physics and modeling of Hall thrusters", *Journal of Applied Physics* **121** 011101 (2017)

[2] Taccogna F and Garriguez L, "Latest progress in Hall thrusters plasma modelling", *Reviews of Modern Plasma Physics*, Springer Singapore, 3 (**1**) (2019)

[3] Lafleur T, Baalrud SD, Chabert P., "Theory for the anomalous electron transport in Hall effect thrusters. I. Insights from particle-in-cell simulations," *Phys. Plasmas* **23**, 053502 (2016)

[4] Katz I, Chaplin VH, Ortega AL, "Particle-in-cell simulations of Hall thruster acceleration and near plume regions", *Phys. Plasmas* **25**, 123504 (2018)

[5] Mikellides IG, Jorns B, Katz I, Ortega AL, "Hall2De Simulations with a First-principles Electron Transport Model Based on the Electron Cyclotron Drift Instability", *52nd AIAA/SAE/ASEE Joint Propulsion Conference*, Salt Lake City, Utah (2016) DOI: 10.2514/6.2016-4618

[6] Reza M, Faraji F, Andreussi T, Andrenucci M "A Model for Turbulence-Induced Electron Transport in Hall Thrusters", IEPC-2017-367, *35th International Electric Propulsion Conference*, Atlanta, Georgia (2017)

[7] Jorns B, "Predictive, data-driven model for the anomalous electron collision frequency in a Hall effect thruster", *Plasma Sources Sci. Technol.* **27** 104007 (2018)

[8] Kaganovich I.D. et al, "Physics of E × B discharges relevant to plasma propulsion and similar technologies", *Phys. Plasmas* **27**, 120601 (2020)

[9] Taccogna F, Longo S, Capitelli M, Schneider R, "Self-similarity in Hall plasma discharges: Applications to particle models", *Phys. Plasmas* **12**, 053502 (2005)

[10] Faraji F, Reza M, Knoll A, "Enhancing one-dimensional particle-in-cell simulations to self-consistently resolve instability-induced electron transport in Hall thrusters", *J. Appl. Phys.* **131**, 193302 (2022)

[11] Reza M, Faraji F, Knoll A, "Resolving multi-dimensional plasma phenomena in Hall thrusters using the reduced-order particle-in-cell scheme", pre-print, 10.21203/rs.3.rs-1693455/v1 (2022)

[12] Bezanson J, Edelman A, Karpinski S, Shah VB, "Julia: A Fresh Approach to Numerical Computing", *SIAM Review*, **59**: 65–98 (2017)

[13] Birdsall CK, Langdon AB, "Plasma Physics via Computer Simulation", CRC Press, 1991

[14] Brieda L, "Plasma Simulations by Example", CRC Press, 2019

[15] Charoy T et al., "2D axial-azimuthal particle-in-cell benchmark for low-temperature partially magnetized plasmas", *Plasma Sources Sci. Technol.* **28** 105010 (2019)

| | |
|---|---|
| [16] | Iserles A, "A First Course in the Numerical Analysis of Differential Equations", 2nd ed., Cambridge Texts in Applied Mathematics (2008) |
| [17] | https://fenicsproject.org/olddocs/dolfin/1.6.0/python/demo/documented/periodic/python/documentation.html, accessed on August 25, 2022. |
| [18] | Charoy T, "Numerical study of electron transport in Hall thrusters", Plasma Physics [physics.plasm-ph], Institut Polytechnique de Paris. English. ⟨NNT: 2020IPPAX046⟩ ⟨tel-02982367⟩ (2020) |
| [19] | Boeuf JP, Garrigues L, "E × B electron drift instability in Hall thrusters: Particle-in-cell simulations vs. theory", *Phys. Plasmas* **25**, 061204 (2018) |
| [20] | Gary SP, Sanderson JJ, "Longitudinal waves in a perpendicular collisionless plasma shock: I. Cold ions", *Journal of Plasma Physics*, Volume 4, Issue 4, pp. 739 – 751 (1970) |
| [21] | Forslund DW, Morse RL, Nielson CW, Electron Cyclotron Drift Instability and Turbulence", *Phys. of Fluids* **15**, 1303 (1972) |
| [22] | Lampe M, Manheimer WM, McBride JB, Orens JH, Shanny R, and Sudan RN, "Nonlinear development of the beam-cyclotron instability", *Phys. Rev. Lett*. **26**, 1221 (1971) |
| [23] | Lampe M, Manheimer WM, McBride JB, Orens JH, Papadopoulos K, Shanny R, and Sudan RN, "Theory and simulation of the beam cyclotron instability," *Phys. Fluids,* **15**, 662 (1972) |
| [24] | Cavalier J, Lemoine N, Bonhomme G, Tsikata S, Honore C, and Gresillon D, "Hall thruster plasma fluctuations identified as the E3B electron drift instability: Modeling and fitting on experimental data", *Phys. Plasmas* **20**, 082107 (2013) |
| [25] | MATLAB. version 9.3.0 (R2017b). Natick, Massachusetts: The MathWorks Inc. (2010) |
| [26] | Lafleur T, Baalrud SD, Chabert P, "Theory for the anomalous electron transport in Hall effect thrusters. II. Kinetic model", *Phys. Plasmas* **23**, 053503 (2016) |
| [27] | Tavant A. "Plasma-wall interaction and electron transport in Hall Effect Thrusters". Thesis work, Université Paris Saclay, France (2019) |
| [28] | Lafleur T, and Chabert P, "The role of instability-enhanced friction on 'anomalous' electron and ion transport in Hall-effect thrusters", *Plasma Sources Science and Technology* **27** 015003 (2018) |
| [29] | Charoy T, Lafleur T, Tavant A, Chabert P, Bourdon A, "A comparison between kinetic theory and particle-in-cell simulations of anomalous electron transport in E×B plasma discharges", *Phys. Plasmas* **27**, 063510 (2020) |
| [30] | Charoy T, Lafleur T, Laguna Alvarez A, Bourdon A, Chabert P, "The interaction between ion transit-time and electron drift instabilities and their effect on anomalous electron transport in Hall thrusters" *Plasma Sources Sci. Technol.* **30** 065017 (2021) |